\documentclass[aps,prb,showpacs,floatfix,twocolumn,byrevtex]{revtex4-1}
\pdfoutput=1
\usepackage{amstext}
\usepackage{amsfonts}
\usepackage{amssymb}
\usepackage{bm}
\usepackage{color}
\usepackage{dcolumn}
\usepackage{graphicx}
\usepackage{hyperref}
\usepackage{wasysym}

\begin{document}

\preprint{Draft --- not for distribution}

%
%
\title{Determination of the optical properties of \boldmath La$_{2-x}$Ba$_x$CuO$_4$ \unboldmath
 for several dopings, including the anomalous \boldmath $x=1/8$ \unboldmath phase }
\author{C. C. Homes}
\email{homes@bnl.gov}
\author{M. H\"{u}cker}
\author{Q. Li}
\author{Z. J. Xu}
\author{J. S. Wen}
\author{G. D. Gu}
\author{J. M. Tranquada}
\affiliation{Condensed Matter Physics and Materials Science Department,
  Brookhaven National Laboratory, Upton, New York 11973, USA}%
\date{\today}

%
%
\begin{abstract}
The optical properties of single crystals of the high-temperature superconductor
La$_{2-x}$Ba$_x$CuO$_4$ have been measured over a wide frequency and temperature
range for light polarized in the {\em a-b} planes and along the {\em c} axis.
Three different Ba concentrations have been examined, $x=0.095$ with a critical
temperature $T_c=32$~K, $x=0.125$ where the superconductivity is dramatically weakened
with $T_c \simeq 2.4$~K, and $x=0.145$ with $T_c\simeq 24$~K.  The in-plane behavior
of the optical conductivity for these materials at high temperature is
described by a Drude-like response with a scattering rate that decreases with
temperature.  Below $T_c$ in the $x=0.095$ and 0.145 materials there is a clear
signature of the formation of a superconducting state in the optical properties
allowing the superfluid density ($\rho_{s0}$) and the penetration depth to be
determined.
In the anomalous 1/8 phase, some spectral weight shifts from lower to higher
frequency ($\gtrsim 300$~cm$^{-1}$) on cooling below the spin-ordering temperature
$T_{\rm so} \simeq 42$~K, associated with the onset of spin-stripe order;
we discuss alternative interpretations in terms of a conventional density-wave
gap versus the response to pair-density-wave superconductivity.
%
%
The two dopings for which a superconducting response is observed both fall on the
universal scaling line $\rho_{s0}/8 \simeq 4.4 \sigma_{dc}T_c$, which is consistent
with the observation of strong dissipation within the {\em a-b} planes.
The optical properties for light polarized along the {\em c} axis reveal an insulating
character dominated by lattice vibrations, superimposed on a weak electronic background.
No Josephson plasma edge is observed in the low-frequency reflectance along the {\em c} axis for
$x=1/8$; however, sharp plasma edges are observed for $x=0.095$ and 0.145 below $T_c$.

\end{abstract}
%
%
%
%
%
%
%
%
\maketitle

\section{Introduction}
\label{sec:intro}
The discovery of superconductivity at elevated temperatures in the
copper-oxide materials a quarter of a century ago sparked an intense
effort to understand the mechanism of the superconductivity in these
compounds.  Despite the wealth of information from the accumulation
of hundreds of thousands of scientific papers, there is still no consensus
on the pairing mechanism.  While superconductivity was originally observed
in the single-layer La$_{2-x}$Ba$_x$CuO$_4$ system,\cite{bednorz86} single
crystals of this material proved difficult to grow and attention quickly
shifted to the related La$_{2-x}$Sr$_x$CuO$_4$ materials where large
single crystals were available. Further investigations into systems with
more than two copper-oxygen layers in the unit cell led to rapid
increases in the critical temperature ($T_c$), with the current
maximum $T_c \simeq 135$~K observed at ambient pressure in the
HgBa$_2$Ca$_2$Cu$_3$O$_{8+\delta}$ system.\cite{chu93}  The physical
properties of this class of correlated electron materials and the
nature of the superconductivity have been described in numerous review articles.\cite{ginsberg-book,harlingen95,timusk99,orenstein00,damascelli03,
basov05,lee06,basov11a}
While most high-temperature superconductors display a doping-dependent
dome-shaped superconducting phase boundary,\cite{takagi89,presland91,
tallon95,broun08} an anomalous weakening of the superconductivity is
observed at a doping of $p=1/8$ per copper atom in the related
La$_{2-x}$Sr$_x$CuO$_4$  system,\cite{moodenbaugh88,yamada92} and more recently
in the Bi$_2$Sr$_2$CaCu$_2$O$_{8+\delta}$ and YBa$_2$Cu$_3$O$_{7-\delta}$
materials.\cite{akoshima98,liang06}  This so-called ``$1/8$ anomaly''
increases dramatically in La$_{2-x}$Ba$_x$CuO$_4$ for $x=1/8$ and
results in the almost total destruction of the bulk three-dimensional
(3D) superconductivity.\cite{moodenbaugh88}  There is also a
structural transition over much of the phase diagram from a
low-temperature orthorhombic (LTO) to a low-temperature tetragonal (LTT)
symmetry\cite{axe89,kumagai91,billinge93,katano93} that is not observed
in pure La$_{2-x}$Sr$_x$CuO$_4$ crystals; the LTO$\rightarrow$LTT transition
is also accompanied by the formation of an antiferromagnetically (AFM)
ordered state.\cite{luke91,luke97,arai03}

Recently, large single crystals of La$_{2-x}$Ba$_x$CuO$_4$ have become
available that have allowed a much more detailed investigation of the
electronic\cite{homes06,xu07,li07,tranquada08,wen08,kim08,kim09,wakimoto09,wen10,adachi11} and
structural\cite{fujita04,tranquada04,abbamonte05,reznik06,zhao07,
dunsiger08a,dunsiger08b,zimmerman08,hucker08,hucker10,hucker11,wilk11}
properties of the 1/8 phase.  With the onset of the LTT phase both long-range
charge-stripe order, followed by spin-stripe order at a slightly lower
temperature, are observed.\cite{tranquada04}
The question of what role stripes play in the formation of a superconducting
state in the cuprate materials is the subject of some debate;\cite{zaanen89,
tranquada95,kivelson03,vojta09a,vojta09b,ekino11} however, in this case it appears
that the formation of static charge-stripe order has the effect of electronically
decoupling the copper-oxygen planes, significantly increasing the resistance
along the {\em c} axis, and frustrating the formation of a bulk superconducting
ground state.\cite{berg07,ding08}
The 1/8 phase exhibits a pseudogap\cite{timusk99} in the LTO phase,\cite{valla06,he09}
similar to what is observed in the underdoped cuprates.\cite{kanigel06} The
remaining Fermi arc (or pocket) in the nodal region becomes gapped in the
LTT phase close to the spin-ordering temperature.  The gap in the nodal
region is momentum-dependent with a {\em d}-wave form, $\Delta(\phi) =
\Delta_0 \cos{2\phi}$, where $\Delta_0$ is the gap maximum and $\phi$
is a Fermi surface angle, similar to the form of the superconducting gap
observed in other cuprate superconductors.\cite{kondo07,hufner08} It has
been proposed that this is a superconducting gap, but that
the pairs lack the necessary phase coherence required to form a
condensate.\cite{li07,tranquada08}

The optical properties of a material can provide important information
about the nature of the lattice vibrations and the electronic transport,
including the formation of charge and superconducting energy gaps.
The role of charge-stripe and spin-stripe ordering in the 1/8 phase is of
particular interest.  Previous optical investigations of stripe correlations
in an analogous phase of La$_{1.88-y}$Nd$_y$Sr$_{0.12}$CuO$_4$,\cite{tajima99,tajima01,dumm02}
and in La$_{2-x}$Sr$_x$CuO$_4$,\cite{dumm02,lucarelli03a,padilla05} have
resulted in some disagreement over the association of certain infrared
spectral features with charge order.\cite{lucarelli03b,tajima03}

%
%
In this work we determine the temperature dependence of the optical
properties of single crystals of La$_{2-x}$Ba$_x$CuO$_4$ over a wide
frequency range for three different dopings, $x=0.095$, 0.125 and
0.145 with $T_c$'s of approximately 32, 2.4 and 24~K, respectively, for
light polarized in the conducting {\em a-b} planes, and along the poorly
conducting {\em c} axis.
The optical conductivity in the {\em a-b} planes for all three materials
show a Drude-like metallic behavior for the free-carriers at room temperature
with a scattering rate that decreases with temperature; for the superconducting
$x=0.095$ and 0.145 samples there is a characteristic suppression
of the low-frequency conductivity below $T_c$, allowing the superfluid
density $\rho_{s0}$ to be determined from this so-called missing
spectral weight.\cite{tinkham}  In contrast, in a refinement of earlier
optical work,\cite{homes06,schafgans10} the 1/8 phase shows a dramatic
narrowing of the free-carrier response below $T_{\rm so}$ accompanied
by the transfer of some spectral weight to both high and low frequencies.
This transfer of spectral weight is consistent with the formation of a
momentum-dependent gap due to the formation of charge-stripe and
spin-stripe order resulting in a nodal metal.  The loss of low-frequency
spectral weight also mimics the missing spectral weight observed
in the superconducting materials, and it is possible that the transfer
of spectral weight is associated with the two-dimensional (2D)
fluctuating superconducting state prior to the onset of bulk 3D
superconductivity.\cite{tranquada08}
The two dopings for which superconductivity is observed fall on the universal
scaling line, $\rho_{s0}/8 \simeq 4.4\,\sigma_{dc}\, T_c$ (where $\sigma_{dc}$
is measured just above $T_c$),\cite{homes04,homes05a,homes05b,homes09} which is
consistent with the observation that there is substantial dissipation in the
normal state of these materials.

The optical properties along the poorly-conducting {\em c} axis
are dominated by the infrared-active lattice modes superimposed
on a weak electronic background.  In the superconducting $x=0.095$ and 0.145
samples, sharp plasma edges in the low-frequency reflectance develop below
$T_c$ due to Josephson coupling between the planes.  No Josephson plasma
edge, or any indication of metallic behavior, is observed for the 1/8 phase
down to the lowest measured temperature, despite the dramatic reduction of
the {\em c}-axis resistivity below $T_{\rm so}$.

%
%
\begin{figure}[t]
\includegraphics[width=0.95\columnwidth]{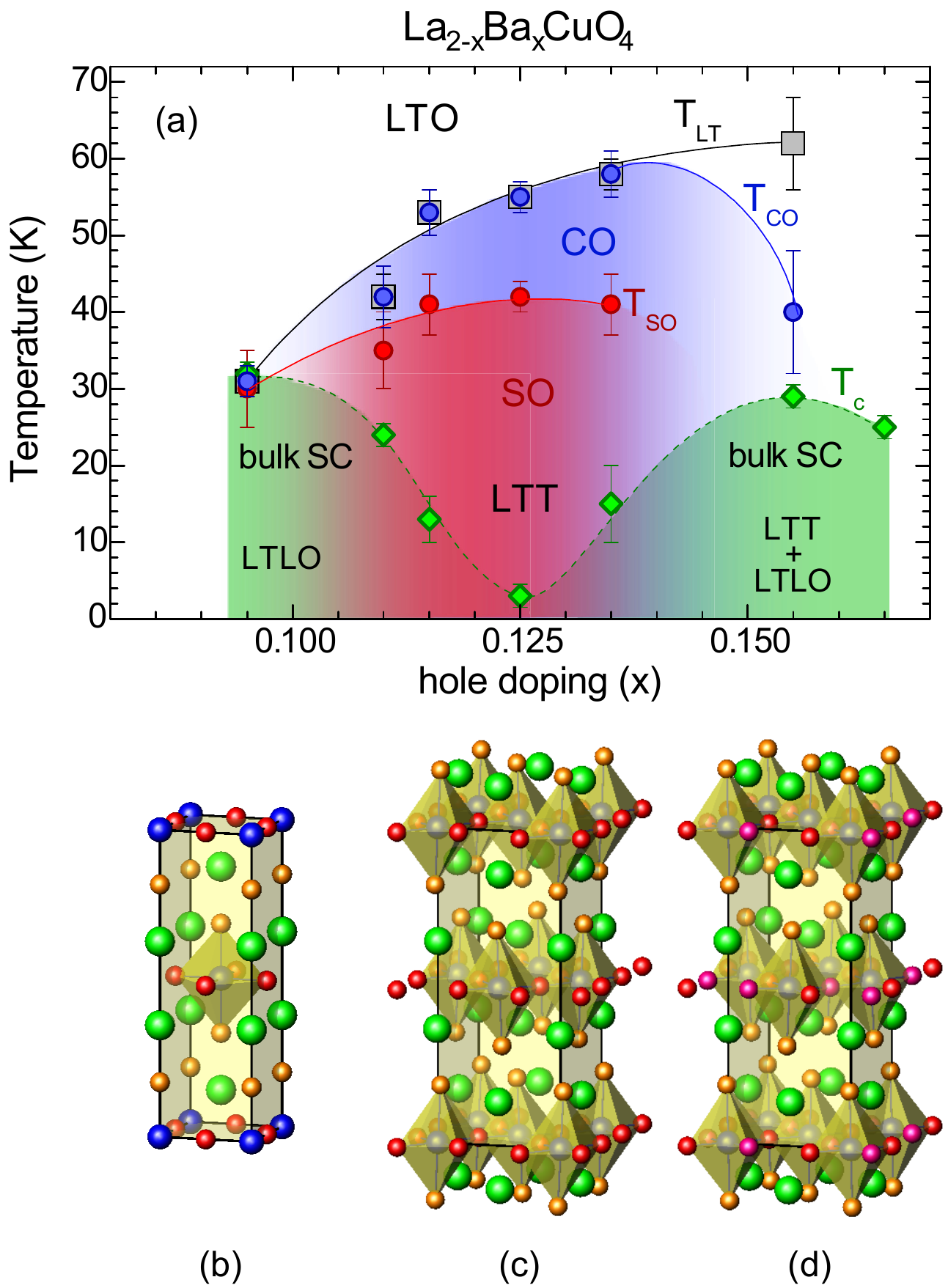}
\caption{
(a) Temperature vs hole doping phase diagram of La$_{2-x}$Ba$_{x}$CuO$_4$
single crystals (reproduced from Ref.~\onlinecite{hucker11}).  Onset
temperatures: $T_c$ for bulk superconductivity (SC), $T_{\rm co}$
for charge stripe order (CO), $T_{\rm so}$ for spin stripe order
(SO), and $T_{\rm LT}$ for the low-temperature structural phases
LTT and LTLO.
The unit cell of La$_{2-x}$Ba$_{x}$CuO$_4$ is shown in the
(b) high-temperature tetragonal (HTT), (c) low-temperature
orthorhombic (LTO), and (d) low-temperature tetragonal (LTT)
phases.}
\label{fig:phase}
\end{figure}

%
%
\section{Experiment}
\label{sec:exper}
Single crystals of La$_{2-x}$Ba$_x$CuO$_4$ were grown using the
traveling-solvent floating-zone method for a series of three
nominal Ba concentrations, $x_0=0.095$, 0.125 and 0.155.
The $x_0=0.095$ and 0.155 compositions are superconducting with
transitions at $T_c \simeq 32$ and 24~K; the superconductivity in
the $x=0.125$ sample is strongly suppressed with $T_c \lesssim 2.4$~K
(all values for $T_c$ were determined by magnetic susceptibility).
The $T_c$ for the nominal $x_0=0.155$ crystal studied here (and in
Ref.~\onlinecite{schafgans10}) is substantially lower than that of the
piece (from the same growth) characterized in Ref.~\onlinecite{hucker11}.
Interpreting the discrepancy in $T_c$ as a difference in Ba concentration,
we estimate that the present sample corresponds to $x=0.145\pm0.01$, and we
refer to it as $x=0.145$ throughout the rest of the paper.
%
%

%
%
\begin{figure}[t]
\includegraphics[width=0.95\columnwidth]{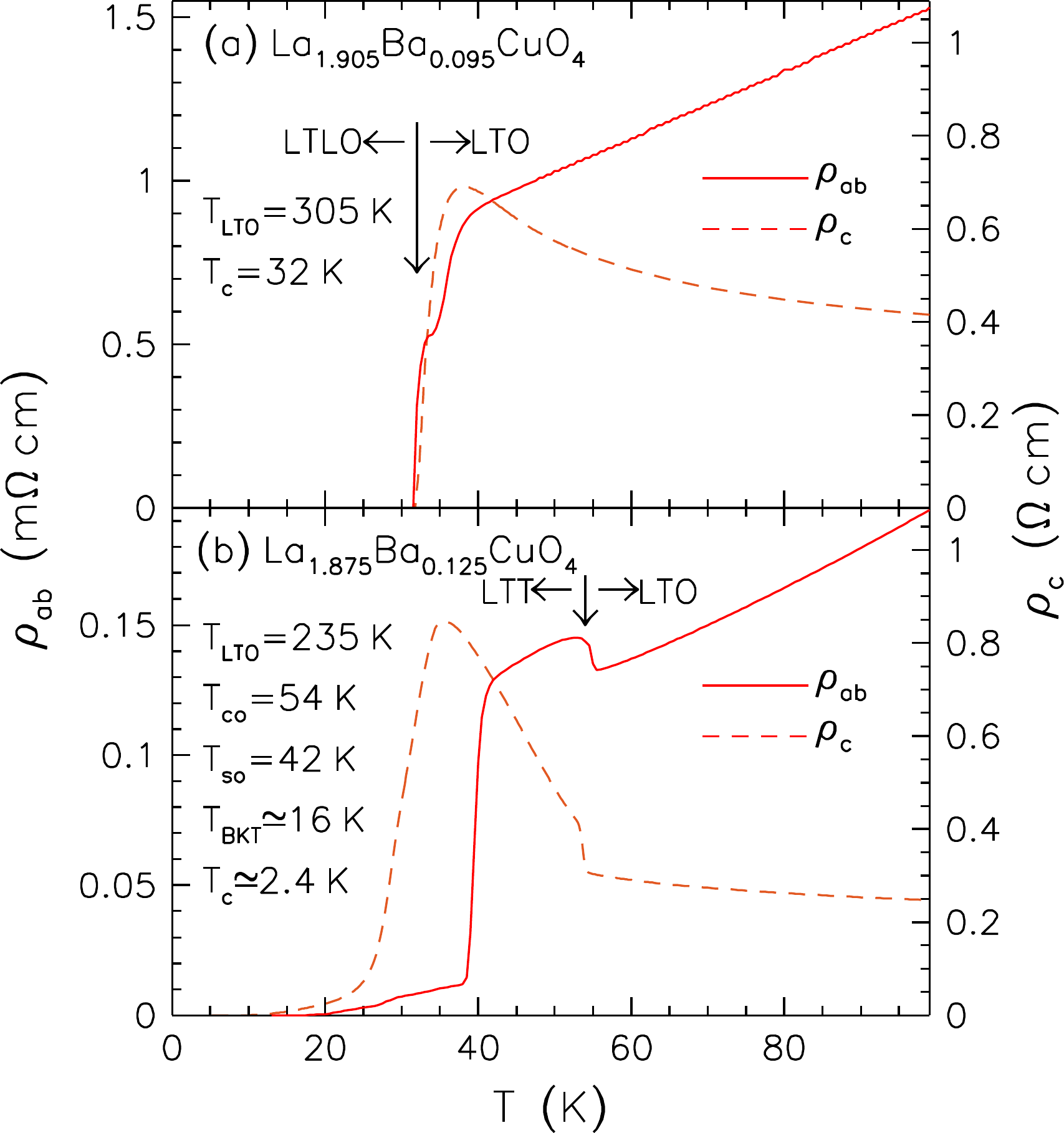}
\caption{
(a) The temperature dependence of the {\em ab}-plane (solid line) and
{\em c}-axis (dashed line) resistivity for $x=0.095$ (Ref.~\onlinecite{wen10}).
In the normal state $\rho_{ab}$ decreases linearly with decreasing temperature,
while $\rho_c$ is observed to increase until just above $T_c$, below which both
decrease dramatically and a weak anomaly is observed in $\rho_{ab}$.
(b) The {\em ab}-plane and {\em c}-axis  resistivity for the $x=0.125$ material
(Refs.~\onlinecite{li07,tranquada08}).  Weak anomalies are observed in $\rho_{ab}$
and $\rho_c$ at $T_{\rm co}$; $\rho_{ab}$ continues to decrease until
$T_{\rm so}$, below which it falls dramatically before once again adopting
a linear temperature dependence.  Along the {\em c} axis, $\rho_c$ continues
to increase below $T_{\rm co}$ until just below $T_{\rm so}$ where it begins
to decrease quickly until $\simeq 25$~K, below which it decreases less rapidly.
Note different scales for $\rho_{ab}$.}
\label{fig:trans}
\end{figure}

%
%
The phase diagram of La$_{2-x}$Ba$_x$CuO$_4$ shown in Fig.~\ref{fig:phase}(a)
is considerably more sophisticated than that of the related La$_{2-x}$Sr$_x$CuO$_4$
material.\cite{hucker11}
Above room temperature, this material is in the $I4/mmm$ tetragonal
(HTT) phase [Fig.~\ref{fig:phase}(b)]. With decreasing
temperature it undergoes an orthorhombic distortion into a $Bmab$
(LTO) phase [Fig.~\ref{fig:phase}(c)] which may be described
as a tilting of the CuO$_6$ octahedra towards the La ions
leading to a roughly $\sqrt{2} \times \sqrt{2}$ larger basal plane
rotated by $45^\circ$.  The HTT$\rightarrow$LTO transition occurs
at $T_{\rm LTO} \simeq 305$, 235 and 190~K for the $x=0.095$, 0.125
and 0.145 materials, respectively.\cite{hucker11}
At dopings close to $x=0.125$ the tilt of the octahedra moves between
the La ions and is accompanied by a weak in-plane distortion resulting
in a low-temperature tetragonal (LTT) phase $P4_2/ncm$ [Fig.~\ref{fig:phase}(d)];
this phase is associated with the development of charge-stripe and spin-stripe
order.\cite{barisic90}
In the $x=0.095$ crystal close to $T_{\rm co} \simeq T_{\rm so}  \simeq T_c$
the LTO structure undergoes a weakening of the orthorhombic strain,
leading to a ``low-temperature less orthorhombic'' (LTLO) $Pccn$ phase
at very low temperature.\cite{wen10}   In the $x=0.145$ material the LTO$\rightarrow$LTT
transition occurs at $\simeq 62$~K, but charge order does not develop
until $T_{\rm co}\simeq 56$~K and the spin order is likely very
weak and is not observed; at low temperature the structure is thought
to be a mixture of LTT$+$LTLO.

%
%
\begin{figure}[t]
\includegraphics[width=0.85\columnwidth]{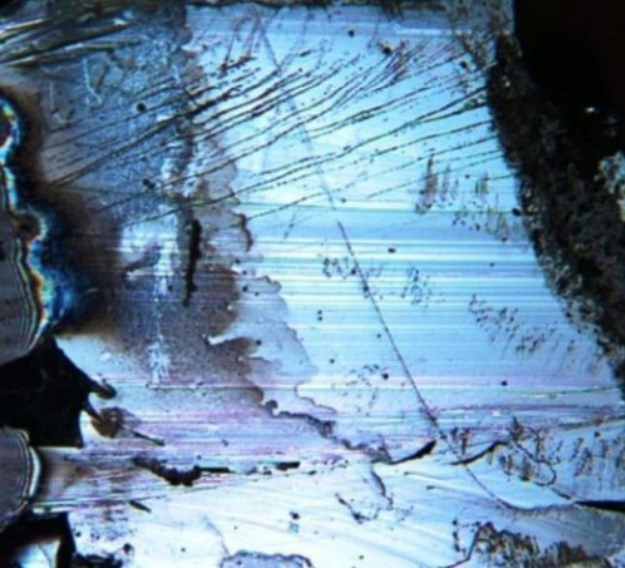}
\caption{A photomicrograph of the cleaved {\em ab}-plane surface
of La$_{1.875}$Ba$_{0.125}$CuO$_4$ revealing a number of striations
and cleavage steps; the total area shown is roughly
$2\,{\rm mm} \times 2\,{\rm mm}$.}
\label{fig:cleave}
\end{figure}

%
%
%
The {\em ab}-plane and {\em c}-axis resistivity of the $x=0.095$ material is
shown close to $T_c \simeq 32$~K in Fig.~\ref{fig:trans}(a) (Ref.~\onlinecite{wen10}).
In the normal state $\rho_{ab}$ displays a characteristic linear
decrease with temperature, until just above $T_c$ where a weak anomaly
is observed that is associated with the onset of the structural transition
previously described.\cite{hucker11,wen11}  In contrast, in the normal state $\rho_c$
increases with decreasing temperature, leading to a resistivity anisotropy
$\rho_c/\rho_{ab}$ that continues to increase until just above $T_c$.
%
%
The {\em ab}-plane and {\em c} axis resistivity of the $x=0.125$ material
is shown in Fig.~\ref{fig:trans}(b) (Refs.~\onlinecite{li07,tranquada08}).
In the normal state $\rho_{ab}$ decreases with temperature in an approximately
linear fashion until a small jump is observed at the charge-ordering temperature
($T_{\rm co}\simeq 54$~K), followed by a dramatic decrease at the
spin-ordering temperature ($T_{\rm so}\simeq 42$~K); this latter
transition is extremely sensitive to the applied magnetic field.\cite{li07}
Below $T_{\rm so}$ a 2D fluctuating superconducting state is thought to form
and the resistivity continues to decrease exponentially; at the same
time a strong diamagnetic response is observed.\cite{tranquada08}
This material is conjectured to undergo a Berezinskii-Kosterlitz-Thouless
transition\cite{berezinskii72,kosterlitz73} at $T_{\rm BKT} \simeq 16$~K,
below this temperature the resistivity is effectively zero.\cite{li07}
A true bulk (3D) superconducting transition is finally achieved at $T_c \simeq 2.4$~K.
%
%
In the normal state $\rho_c$ shows little change
with decreasing temperature until $T_{\rm co}$ where, similar to $\rho_{ab}$,
a small jump is observed; however, $\rho_c$ continues to increase indicating a
dramatic increase in the resistivity anisotropy until slightly below $T_{\rm so}$,
where $\rho_c$ begins to decrease quickly, reaching a negligible value below
$\simeq 6$~K.\cite{tranquada08}

\begin{figure}[t]
\includegraphics[width=0.95\columnwidth]{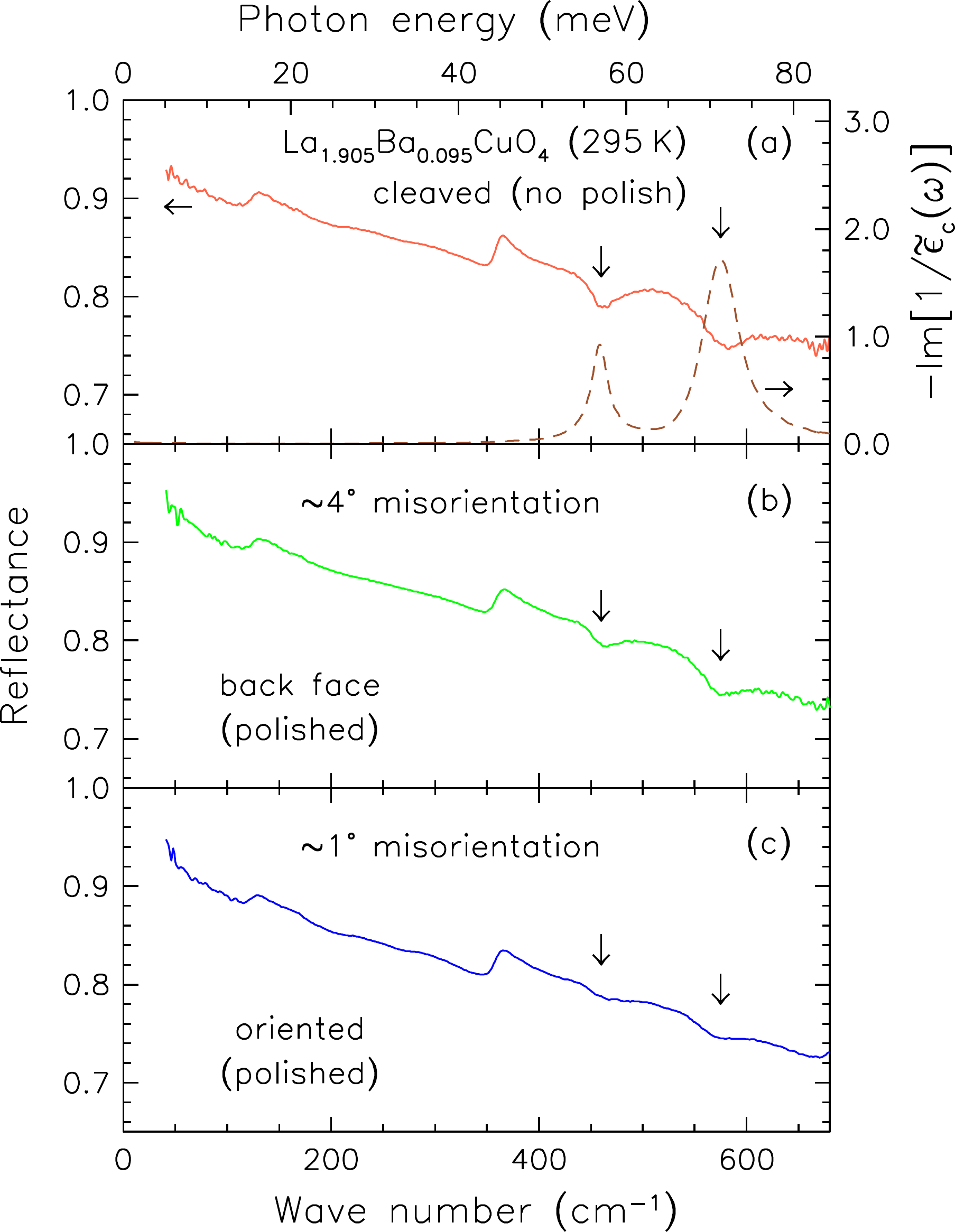}
\caption{The reflectance of La$_{1.905}$Ba$_{0.095}$CuO$_4$
at 295~K in the far-infrared region for light polarized
in the {\em a-b} planes.
(a) The reflectance from a cleaved surface.  There are two
prominent antiresonances at $\simeq 460$ and 570~cm$^{-1}$
which are associated with {\em c}-axis longitudinal optic
modes that become infrared-active due to a surface
misorientation.  This effect can be seen in the {\em c}-axis
loss function (dashed line).
(b) The reflectance from the polished surface of the same
sample with a measured surface misorientation of $4^\circ$;
the antiresonances are still prominent.
(c) The reflectance from the polished surface with a
measured surface misorientation of $\lesssim 1^\circ$;
the antiresonances are substantially weaker.}
\label{fig:faces}
\end{figure}

%
%
It is possible to cleave some of these materials, resulting in
mirror-like {\em ab}-plane surfaces.  However, in  most cases
cleaving yields a ``terraced'' surface consisting of numerous steps with
exposed {\em c}-axis faces, shown in Fig.~\ref{fig:cleave}.
In Fig.~\ref{fig:faces}(a) the absolute reflectance of a
freshly-cleaved {\em ab}-plane surface for the $x=0.095$ material
has been measured at room temperature at a near-normal angle of
incidence over the infrared region using an {\em in situ}
evaporation technique.\cite{homes93a,homes07b}
In addition to the Hagen-Rubens $R\propto 1-\sqrt{\omega}$
behavior expected for a metallic system and the two normally
infrared-active vibrations at about 120 and 360~cm$^{-1}$, there
are two strong antiresonances at $\simeq 460$ and 575~cm$^{-1}$
associated with the longitudinal optic modes of the infrared-active
{\em c}-axis modes (Sec.~IIIB).  These features are absent from
naturally-grown crystal faces, but are observed whenever a surface
has been cut and polished, or as illustrated in the photomicrograph
shown in Fig.~\ref{fig:cleave}, when a cleaved surface has numerous
steps.  This phenomenon has been documented extensively in the literature
and is attributed to the misorientation of the {\em ab}-plane
surface.\cite{reedyk92,kim94,startseva99b,tsvetkov99,tajima05,homes07a}

Clearly, it is desirable to minimize these extrinsic features while
preserving the intrinsic behavior of the sample.  Pursuing this goal,
the opposite face was polished using a series of successively finer
diamond grits using kerosene as a suspension agent, with a final lap
using a 0.1~$\mu$m diamond paste.  The infrared reflectance shown in
Fig.~\ref{fig:faces}(b) is quite similar to the reflectance of the
cleaved surface, and shows that the antiresonance features are still
present.  It was subsequently determined from x-ray Laue
diffraction that this surface was misoriented by $\simeq 4^\circ$.
The sample was reoriented using the Laue camera to better than $1^\circ$
and then repolished.  The resulting reflectance in Fig.~\ref{fig:faces}(c)
indicates that the antiresonance features have been minimized. Thus, while
cleaved surfaces are often desirable, in those cases where strong
antiresonances are observed, determining the correct orientation and
careful polishing may remove most of these unwanted spectral features.

%
%
The reflectance of single crystal La$_{2-x}$Ba$_x$CuO$_4$ for
$x=0.095$, 0.125 and 0.145 has been measured at a near-normal
angle of incidence for light polarized in the {\em a-b} planes,
as well as along the {\em c} axis, over a wide frequency range
from $\simeq 2$~meV to over 3~eV ($\simeq 20$ to over
24,000~cm$^{-1}$) at a variety of temperatures using the previously
described overcoating technique.  The {\em ab}-plane reflectance
of the $x=0.125$ sample was measured from a cleaved surface,
while for the other dopings the surface was oriented to better than
$1^\circ$ and polished.
The {\em c}-axis was revealed by cutting the {\em a-b} face at $90^\circ$,
which was subsequently polished.  The {\em c}-axis direction was determined
by using an infrared polarizer to check the modulation of the infrared
signal, with the minima corresponding to the {\em c}-axis direction.
%

%
%

The reflectance is a complex quantity consisting of an amplitude and
a phase, $\tilde{r} = \sqrt{R}\exp{(i\theta)}$.  Because only the
amplitude $R = \tilde{r}\,\tilde{r}^\ast$ is measured in this
experiment it is often difficult to decipher how the reflectance is
related to the real and imaginary optical properties.
Thus, the complex optical properties have been calculated from a
Kramers-Kronig analysis of the reflectance.\cite{dressel-book}  The
Kramers-Kronig transform requires that the reflectance be determined
for all frequencies, thus extrapolations must be supplied in the
$\omega \rightarrow 0, \infty$ limits.  At low frequency, for light
polarized in the {\em a-b} planes, above $T_c$ in the normal state the
metallic Hagen-Rubens form for the reflectance $R(\omega) \propto
1-\sqrt{\omega}$ is employed.  Below $T_c$ in the superconducting
state $R(\omega) \propto 1-\omega^4$; however, it should be noted
that when the reflectance is close to unity the analysis is not
sensitive upon the choice of low-frequency extrapolation.
For light polarized along the {\em c} axis, at low frequency the
Hagen-Rubens form is used in the normal state, while below $T_c$ we
again use $R(\omega) \propto 1 - \omega^4$.
For all polarizations the reflectance is assumed to be constant above
the highest measured frequency point up to $\simeq 1 \times 10^5$~cm$^{-1}$,
above which a free electron gas asymptotic reflectance extrapolation
$R(\omega) \propto 1/\omega^4$ is employed.\cite{wooten}

\vfil

%
%
\section{Results and Discussion}

\subsection{{\em a-b} plane}
\begin{figure}[t]
\includegraphics[width=0.95\columnwidth]{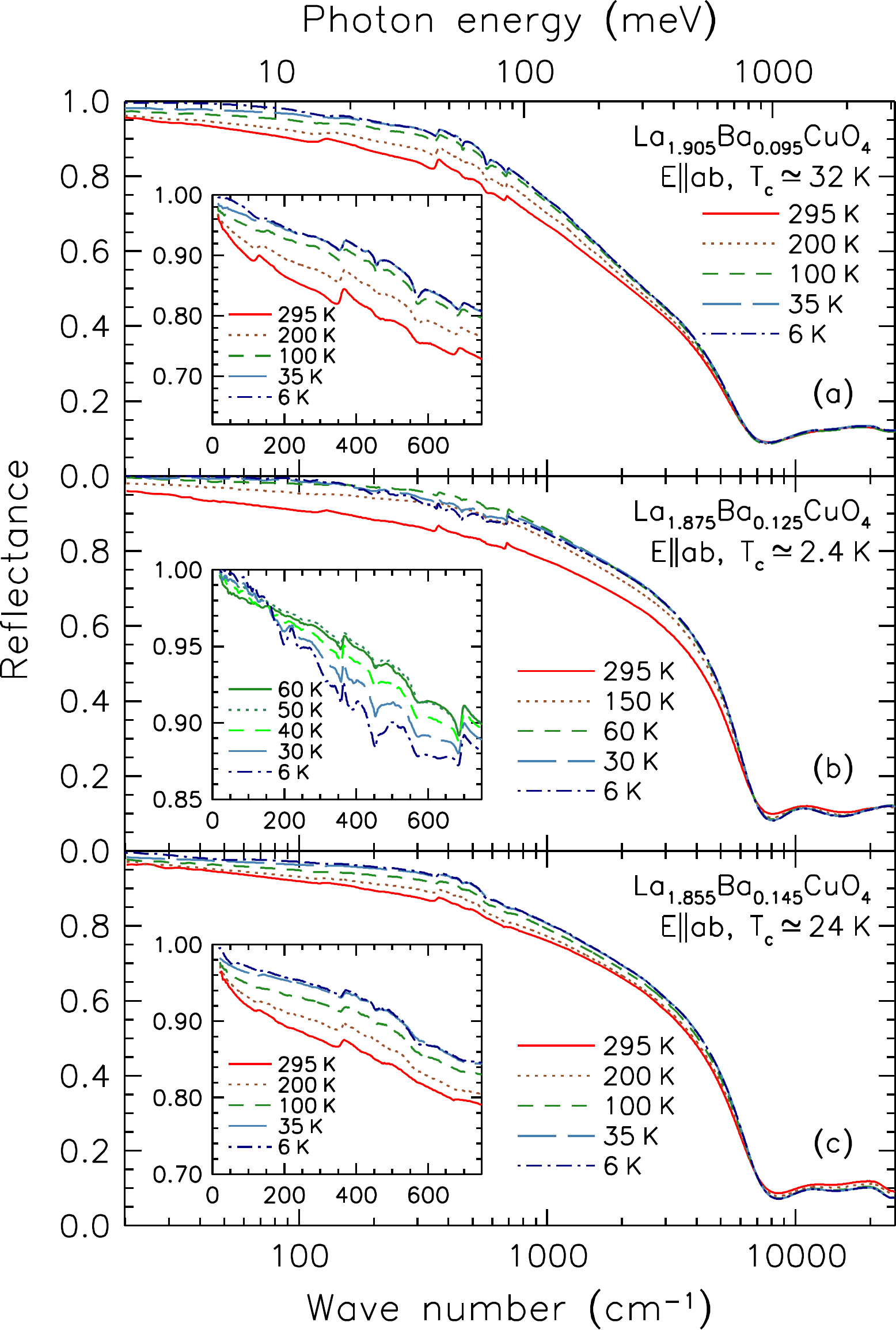}
\caption{The temperature dependence of the reflectance of La$_{2-x}$Ba$_x$CuO$_4$
for light polarized in the {\em a-b} planes over a wide frequency range for:
(a) a polished sample with $x=0.095$, (b) a cleaved sample for $x=0.125$, and
(c) a polished sample with $x=0.145$.
Insets: The temperature dependence of the reflectance  in the far-infrared
region.}
\label{fig:reflec}
\end{figure}

The temperature dependence of the reflectance of La$_{2-x}$Ba$_x$CuO$_4$
for light polarized in the {\em a-b} planes is shown over a wide frequency
range for $x=0.095$, 0.125 and 0.145 in Figs.~\ref{fig:reflec}(a), (b) and
(c), respectively; the insets show the temperature dependence of the
reflectance in the far-infrared region.

%
%

At room temperature the reflectance curves for all three dopings
are similar in that the reflectance at low frequency is high and
decreases with frequency until a plasma edge in the reflectance is
encountered at about 1~eV.  The reflectance in the infrared region
is observed to increase with electronic doping.
In the three samples the temperature dependence of the reflectance
all show a similar increase in the low-frequency reflectance with
decreasing temperature.
In the superconducting state the $x=0.095$ material displays an
abrupt increase in the low-frequency reflectance, while the changes
in the $x=0.145$ material below $T_c$ are not as pronounced.
At low temperature the reflectance of the $x=0.125$ sample is
dramatically different; below $\simeq 200$~cm$^{-1}$ the reflectance
increases down to the lowest measured temperature, but close to
or below $T_{\rm so}$ the reflectance in the $200 - 2000$~cm$^{-1}$
region actually decreases with temperature,\cite{homes06} as shown in
the inset of Fig.~\ref{fig:reflec}(b).  While the reflectance contains
a great deal of information about the electronic transport and structure
in a material, the complicated behavior observed in the $x=0.125$ material
makes it difficult to interpret.  The determination of the optical
conductivity from a Kramers-Kronig analysis of the reflectance is more
revealing.

%
%
\subsubsection{$x=0.095$}
\begin{figure}[t]
\includegraphics[width=0.95\columnwidth]{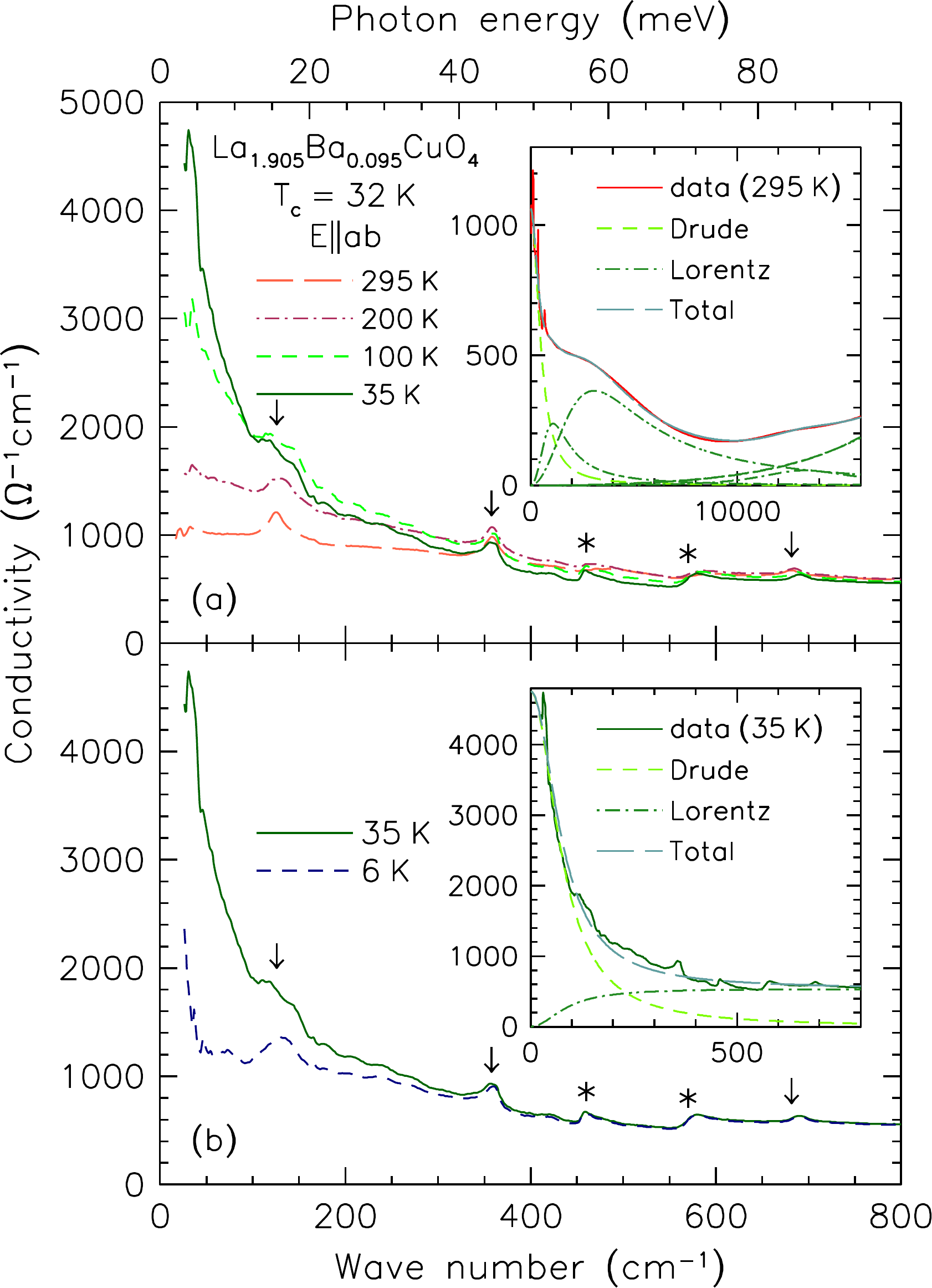}
\caption{The real part of the optical conductivity of La$_{1.905}$Ba$_{0.095}$CuO$_4$
for light polarized in the {\em a-b} planes. (a) The infrared region for
several temperatures above $T_c = 32$~K.  The antiresonances observed in the
reflectance in Fig.~\ref{fig:faces}(c) are indicated by asterisks, while the
in-plane infrared-active lattice modes are indicated by the arrows.
Inset: The Drude-Lorentz fit to the data at 295~K.  The Drude component
(dashed line) and Lorentz oscillators (dash-dot lines) combine (long-dashed
line) to reproduce the data (solid line) quite well.
(b) The optical conductivity just above and well below $T_c$ in the infrared
region illustrating the transfer of spectral weight into the condensate.
Inset: The Drude-Lorentz fit to the data at 35~K.}
\label{fig:sigma1}
\end{figure}

The real part of the optical conductivity for light polarized in the {\em a-b}
planes of La$_{1.905}$Ba$_{0.095}$CuO$_4$ is shown in the infrared region at
temperatures above $T_c$ in Fig.~\ref{fig:sigma1}(a).  At room temperature the
conductivity appears rather broad in the far-infrared region, but as the inset
in this panel shows it is in fact decreasing rather rapidly before giving way
above $\simeq 800$~cm$^{-1}$ to a broad plateau in the mid-infrared; there is
a shoulder at about 4000~cm$^{-1}$ (0.5~eV) above which the conductivity
continues to decrease until about 1~eV, at which point it begins to slowly
rise again.
%
%
Superimposed on the room-temperature conductivity curve are several sharp
resonances at $\simeq 126$, 358 and 682~cm$^{-1}$; these are the normally-active
infrared vibrations which are observable due to the poor screening in this
class of materials.\cite{homes00}  The weak antiresonances observed in the
reflectance in Fig.~\ref{fig:faces}(c) at $\simeq 460$ and 570~cm$^{-1}$
attributed to a surface misorientation manifest themselves as weak
resonances or antiresonances in the conductivity.  In the HTT phase,
the irreducible vibrational representation for the infrared active modes is
$$
  \Gamma_{\rm IR}^{\rm HTT} = 4A_{2u} + 4E_u
$$
where the singly-degenerate $A_{2u}$ modes are active along the {\em c} axis
and the doubly-degenerate $E_u$ modes are active in the {\em a-b} planes.
Below the structural phase transition in the LTO phase the unit cell is larger
and more complicated.  The lowered symmetry results in the removal of
degeneracy and more infrared modes
$$
  \Gamma_{\rm IR}^{\rm LTO} = 4A_u + 6B_{1u} + 7B_{2u} + 4B_{3u}
$$
where the $A_u$ modes are silent, and the singly-degenerate $B_{1u}$,
$B_{2u}$ and $B_{3u}$ modes are active along the {\em c}, {\em b} and
{\em a} axes, respectively.  The additional low-temperature structural
transition to either a LTLO or LTT+LTLO phase represents a further
reduction of symmetry; for the LTT phase the irreducible
vibrational representation becomes
$$
  \Gamma_{\rm IR}^{\rm LTT} = 3A_{1u} + 7A_{2u} + 8B_{1u} +
  3B_{2u} + 12E_{u}
$$
where the $A_{1u}$, $B_{1u}$ and $B_{2u}$ modes are silent; as in the
case of the HTT phase the $A_{2u}$ and $E_u$ modes are active along the
{\em c} axis and the {\em a-b} planes, respectively.

At room temperature La$_{1.905}$Ba$_{0.095}$CuO$_4$ appears to be in the HTT
phase (demonstrated by the {\em c}-axis properties in Sec.~IIIB); of four
possible in-plane modes, only three are clearly identified (the remaining mode
is likely too weak to be observed).  At low temperature the vibrations typically
narrow and increase in frequency; however, the low-frequency mode at
$\simeq 126$~cm$^{-1}$ in Fig.~\ref{fig:sigma1}(a) appears to broaden
anomalously at low temperature, suggesting it may be sensitive to the
charge or spin order.

%
%

%
%
As the temperature decreases the low-frequency spectral weight associated with
the free-carrier component increases rapidly.  The spectral weight is
defined here simply as the area under the conductivity curve over a given
frequency interval: $N(\omega_c, T) = \int_{0^+}^{\omega_c} \sigma_1(\omega, T)
\,d\omega$.  For a metal, in the absence of other excitations, the area under the
entire conductivity curve is the well-known {\em f}-sum rule,\cite{smith}
$\int_{0}^\infty \sigma_1(\omega)\,d\omega = \omega_p^2/8$, where $\omega_p$
is the classical plasma frequency.
The value for the dc resistivity for $x=0.095$ in Fig.~\ref{fig:trans}(a)
just above $T_c$ is in reasonably good agreement with the extrapolated value
for $\sigma_{dc}\equiv \sigma_1(\omega\rightarrow 0)$, although it should be
noted that in this temperature region the resistivity is changing rapidly.
%
%
%
\begin{table}[tb]
\caption{The parameters for the Drude plasma frequency $\omega_{p,D}$ and
scattering rate $1/\tau_D$ determined in the normal state at 100~K.$^a$
The superconducting plasma frequency $\omega_{p,S}$ and effective
penetration depth are determined for $T \ll T_c$.}
\begin{ruledtabular}
\begin{tabular}{c ccc cc}
  $x$   & $T_c$ & $\omega_{p,D}$ & $1/\tau_D$  & $\omega_{p,S}$ & $\lambda_0$ \\
        &  (K)  &  (cm$^{-1}$)  & (cm$^{-1}$) &  (cm$^{-1}$)   & (\AA ) \\
  0.095 & 32          & 4360 & 131 & 3550        & 4480 \\
  0.125 & $\sim{2.4}$ & 6480 & 105 &  $-$        & $-$ \\
  0.145 & 24          & 7360 & 261 & $\sim 2970$ & $\sim 5360$ \\
\end{tabular}
\end{ruledtabular}
\footnotetext[1] {Fits at 100~K; $x=0.125$ interpolated from 60 and 150~K data.}
\label{tab:props}
\end{table}
Angle-resolved photoemission (ARPES) studies of high-temperature superconductors
in general show a single band crossing the Fermi level,\cite{damascelli03}
indicating that the intra-band excitations giving rise to the free-carrier
component have a single origin.  Within this context, it is not unreasonable to
try and model the optical conductivity using a simple Drude-Lorentz model for
the complex dielectric function $\tilde\epsilon(\omega) = \epsilon_1(\omega) +
i\epsilon_2(\omega)$,
\begin{equation}
  \tilde\epsilon(\omega) = \epsilon_\infty
   - {{\omega_{p,D}^2}\over{\omega^2+i\omega/\tau_{D}}}
   + \sum_j {{\Omega_j^2}\over{\omega_j^2 - \omega^2 - i\omega\gamma_j}},
\end{equation}
where $\epsilon_\infty$ is the real part of the dielectric function at high
frequency, $\omega_{p,D}^2 = 4\pi n e^2/m^\ast$ and $1/\tau_{D}$ are the
square of the plasma frequency and scattering rate for the delocalized (Drude)
carriers; $\omega_j$, $\gamma_j$ and $\Omega_j$ are the position, width, and
strength of the $j$th vibration or excitation.  The complex conductivity is
$\tilde\sigma(\omega) = \sigma_1(\omega) + i\sigma_2(\omega) =
-i\omega [\tilde\epsilon(\omega) - \epsilon_\infty ]/4\pi$, yielding the
expression for the real part of the optical conductivity
\begin{equation}
  \sigma_1(\omega) = {1\over{60}}\, {{\omega_{p,D}^2\tau_D} \over
    {1+\omega^2\tau_D^2}} + \sigma_{MIR},
\end{equation}
where the first term is the Drude response and $\sigma_{MIR}$ is the
contribution from the bound (Lorentz) excitations.  (When
$\omega_{p,D}$ and $1/\tau_D$ are in units of cm$^{-1}$, the
conductivity has units of $\Omega^{-1}$cm$^{-1}$.)
The Drude-Lorentz model has been fit to the normal state conductivity using
a non-linear least-squares method at 295 and 35~K.  In addition to the Drude
component, two Lorentz oscillators have been included somewhat arbitrarily at
$\simeq 800$ and 3100~cm$^{-1}$ to reproduce the observed mid-infrared response.\cite{fit}
The fit to the data at 295~K is in good agreement with the experimental data
and is shown over a broad spectral range in the inset of Fig.~\ref{fig:sigma1}(a).
The fit to the data in the normal state at 35~K is quite good and is
shown in the inset in Fig.~\ref{fig:sigma1}(b). The Drude plasma frequency
$\omega_{p,D} \simeq 4360\pm 300$~cm$^{-1}$ does not vary with temperature,
while the scattering rate decreases dramatically with temperature,
$1/\tau_D \simeq 295$, 207, 131 and 75~cm$^{-1}$ at 295, 200, 100
and 35~K, respectively (Table~\ref{tab:props}).

%
%
%
%
For a superconductor with an isotropic energy gap $\Delta$, the
gap in the optical conductivity is $2\Delta$.
In the superconducting state below $T_c$, shown in Fig.~\ref{fig:sigma1}(b)
for $x=0.095$, the dramatic decrease in the low-frequency optical
conductivity below $\sim 200$~cm$^{-1}$ (25~meV) is associated with the
formation of a superconducting energy gap;\cite{tinkham} this result is in
reasonable agreement with the ARPES estimate of $2\Delta_0 \simeq 20$~meV
for the maximum gap value in this material.\cite{valla06,he09}
The ``missing area'' is referred to as the spectral weight of the
condensate $N_c$ and may be calculated  from the Ferrell-Glover-Tinkham
sum rule\cite{ferrell58,tinkham59}
\begin{equation}
  N_c \equiv N(\omega_c, T\simeq T_c) - N(\omega_c, T\ll T_c) = \omega_{p,S}^2/8.
\end{equation}
Here $\omega_{p,S}^2 = 4\pi n_s e^2/m^*$ is the square of the superconducting
plasma frequency, and the superfluid density is simply $\rho_{s0} \equiv \omega_{p,S}^2$;
the cut-off frequency $\omega_c$ is chosen so that the integral converges
smoothly.  The superconducting plasma frequency has also been determined
from the real part of the dielectric function in the low frequency limit
where $\epsilon_1(\omega) = \epsilon_\infty - \omega_{p,S}^2/\omega^2$.
Yet another method of extracting $\omega_{p,S}$ from $\epsilon_1(\omega)$ is
to determine $[-\omega^2\epsilon_1(\omega)]^{1/2}$ well below $T_c$ in
the $\omega\rightarrow 0$ limit.\cite{jiang96} All three techniques yield
$\omega_{p,S} \simeq 3550\pm 200$~cm$^{-1}$, which corresponds to an effective
penetration depth of $\lambda_0 \simeq 4480\pm 270$~\AA .
The fact that less than two-thirds of the free-carriers in the normal state have
condensed ($\omega_{p,S}^2/\omega_{p,D}^2 \lesssim 0.6$) is consistent with the
observation of strong dissipation in the normal state.

%
%
\begin{figure}[t]
\includegraphics[width=0.95\columnwidth]{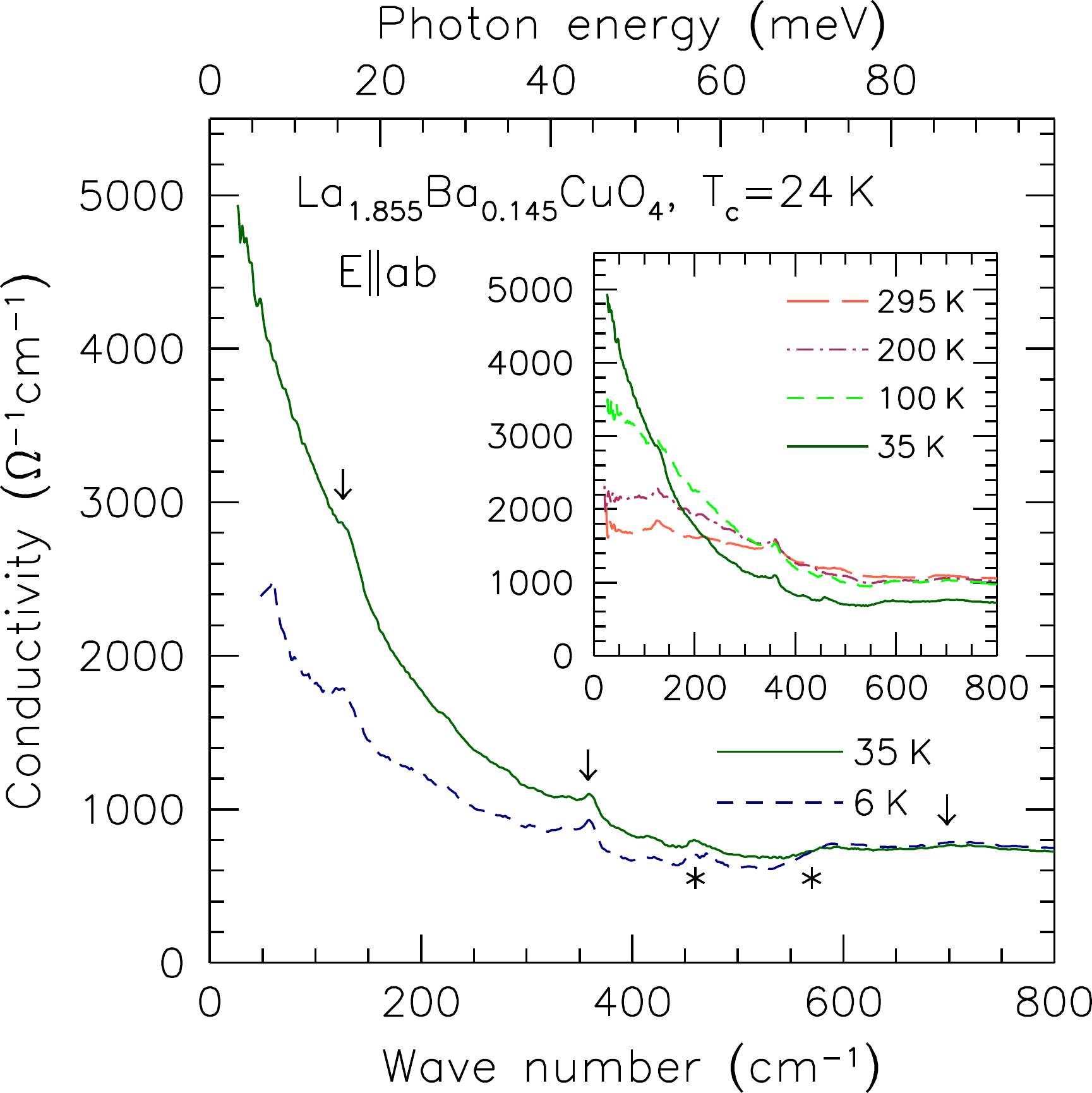}
\caption{The real part of the optical conductivity of La$_{1.855}$Ba$_{0.145}$CuO$_4$
for light polarized in the {\em a-b} planes measured just above and well below
the critical temperature. The asterisks denote structure due to surface
misorientation, while the arrows indicate the in-plane infrared-active lattice
modes.
Inset: The temperature dependence of the optical conductivity in the
normal state.}\label{fig:sigma2}
\end{figure}

%
%
\subsubsection{$x=0.145$}
The temperature dependence of the real part of the optical conductivity for
the $x=0.145$ material is shown in Fig.~\ref{fig:sigma2} at a variety of
temperatures above and below $T_c$.  In the normal state the overall
behavior is not unlike that of the $x=0.095$ material; at low frequency a
Drude-like component narrows rapidly with decreasing temperature but gives
way to a broad, incoherent component above $\sim 60$~meV.  The conductivity
in the normal state can be fit reasonably well using the Drude-Lorentz
model yielding $\omega_{p,D} \simeq 7360$~cm$^{-1}$ and $1/\tau_D \simeq
500$~cm$^{-1}$ at room temperature.  The larger value for $\omega_{p,D}$
denotes the increase in the density of free carriers.  It is interesting
to note that the normal-state scattering rate in the overdoped material is
roughly twice that of the underdoped system; this may be due to electronic
correlations or it may be that the increasing cation disorder in the blocking
layer results in a larger value of $1/\tau_D$ (Table~\ref{tab:props}).
Upon entry into the superconducting state a decrease in the low-frequency
conductivity is once again observed; however, the onset at which this
decrease occurs is rather indistinct and there is considerably more residual
conductivity at low frequency than was observed in the underdoped material.
Using the techniques previously discussed, the superconducting plasma frequency
is estimated to be $\omega_{p,S} \simeq 2970\pm 180$~cm$^{-1}$, corresponding
to a penetration depth of $\lambda_0 \simeq 5360\pm 350$~\AA .  As a result
of the large normal state scattering rate, less than a fifth of the free-carriers
have condensed ($\omega_{p,S}^2/\omega_{p,D}^2 \lesssim 0.2$).

%
%
\subsubsection{$x=0.125$}
\begin{figure}[t]
\includegraphics[width=0.95\columnwidth]{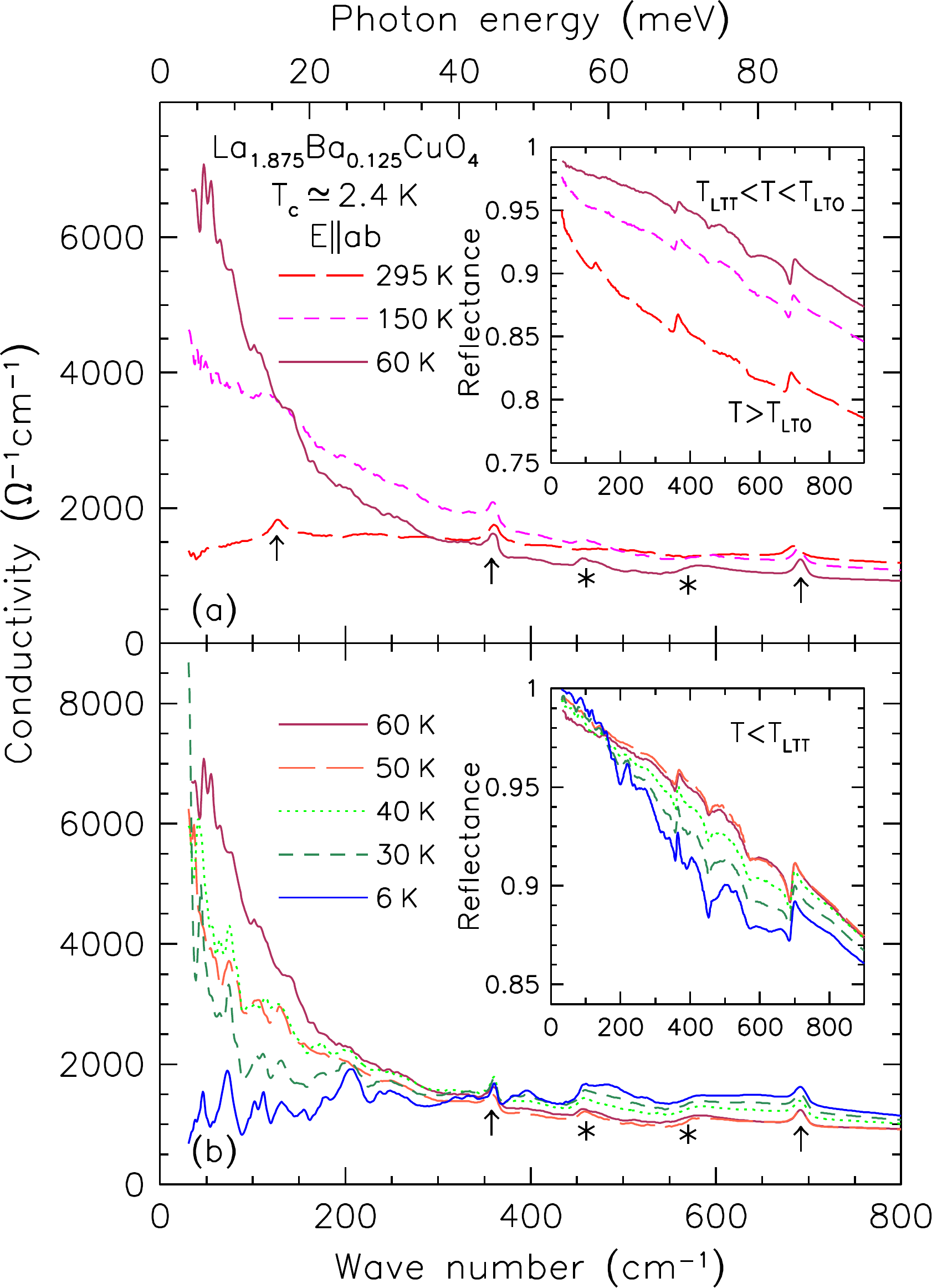}
\caption{(a) The real part of the optical conductivity
La$_{1.875}$Ba$_{0.125}$CuO$_4$ for light polarized
in the {\em a-b} planes at several temperatures above
$T_{\rm co}$ and $T_{\rm so}$. Inset: the infrared reflectance
for the same temperatures.
(b) The conductivity for several temperatures below $T_{\rm co}$
and $T_{\rm so}$ in the same region showing the transfer
and loss of spectral weight occurring mainly below $T_{\rm so}$.
Inset: the infrared reflectance for the same temperatures
contrasting the different responses at low and high frequency.
The asterisks denote structure due to surface misorientation,
while the arrows indicate the in-plane infrared-active lattice
modes.}
\label{fig:sigma3}
\end{figure}

The in-plane reflectance of La$_{2-x}$Ba$_x$CuO$_4$ for $x=0.125$ is shown in
Fig.~\ref{fig:reflec}(b).  While some aspects of the reflectance have been
previously discussed,\cite{homes06} there are several important points that
bear repeating.  As the temperature is lowered, the reflectance is observed
to increase in the far- and mid-infrared regions.  Close to or below the
spin-ordering temperature $T_{\rm so} = 42$~K the reflectance decreases in
the $200 - 2000$~cm$^{-1}$ interval; however, below about 200~cm$^{-1}$ the
reflectance continues to increase.\cite{homes06}
The consequences of this behavior may be seen more clearly in the
conductivity shown for temperatures above $T_{\rm co}$
in Fig.~\ref{fig:sigma3}(a) and mainly below $T_{\rm co}$ in
Fig.~\ref{fig:sigma3}(b).
%
%
At high temperature the optical conductivity can be modeled quite well
using a Drude-Lorentz model; as the temperature is lowered from room
temperature to just above $T_{\rm co}=54$~K, there is a characteristic
narrowing of the Drude-like component and spectral weight is transferred
from high to low frequency.  The plasma frequency $\omega_{p,D}
\simeq 6480\pm 350$~cm$^{-1}$ is relatively constant, while the scattering
rate decreases from $1/\tau_D=439$~cm$^{-1}$ at room temperature to 81~cm$^{-1}$ at
$T_{\rm co}$ (Table~\ref{tab:props}).

In Fig.~\ref{fig:sigma3}(b) for  $T_{\rm so} < T < T_{\rm co}$ the scattering
rate continues to decrease without any obvious loss of spectral weight or
change in $\omega_{p,D}$.  Below $T_{\rm so}$ the Drude-like component in
the conductivity continues to narrow; however, above $\sim 300$~cm$^{-1}$
there is now a noticeable increase in the conductivity, indicating a transfer
of spectral weight from the low-frequency free-carrier component to localized
or gapped excitations,\cite{homes03} the significance of which will be
discussed in Sec. IIIC.
%
%
This is precisely the temperature at which the
infrared reflectance, shown in the inset of Fig.~\ref{fig:sigma3}(b),
begins to decrease above $\simeq 200$~cm$^{-1}$, while still increasing
at low frequency.  Below $\simeq 30$~K little spectral weight is
being transferred to high energy, instead the decrease in the low-frequency
spectral weight is now associated almost entirely with the increasingly
narrow free-carrier response, such as might be expected in a nodal
metal.\cite{ando01,sutherland03,lee05}. Attempts to fit the free-carrier
component with a Drude-model yield a rapidly decreasing value for
$\omega_{p,D}$ for $T < T_{\rm so}$, but at the same time $1/\tau_D$
is also decreasing dramatically.\cite{homes06} We note that in the
Drude model $\sigma_{dc} = \omega_{p,D}^2\tau_D/60$, so that the
net effect is the rapidly decreasing resistivity observed in
Fig.~\ref{fig:trans}(b).
At $\sim 6$~K the free-carrier response is now so narrow it may no longer
be accurately observed, giving way to a broad, incoherent background
conductivity.  It is difficult to distinguish this behavior from the
phenomenon of missing spectral weight in the superconducting state.
%
%

%
%
\subsubsection{Scattering rate}
\begin{figure}[t]
\includegraphics[width=1.00\columnwidth]{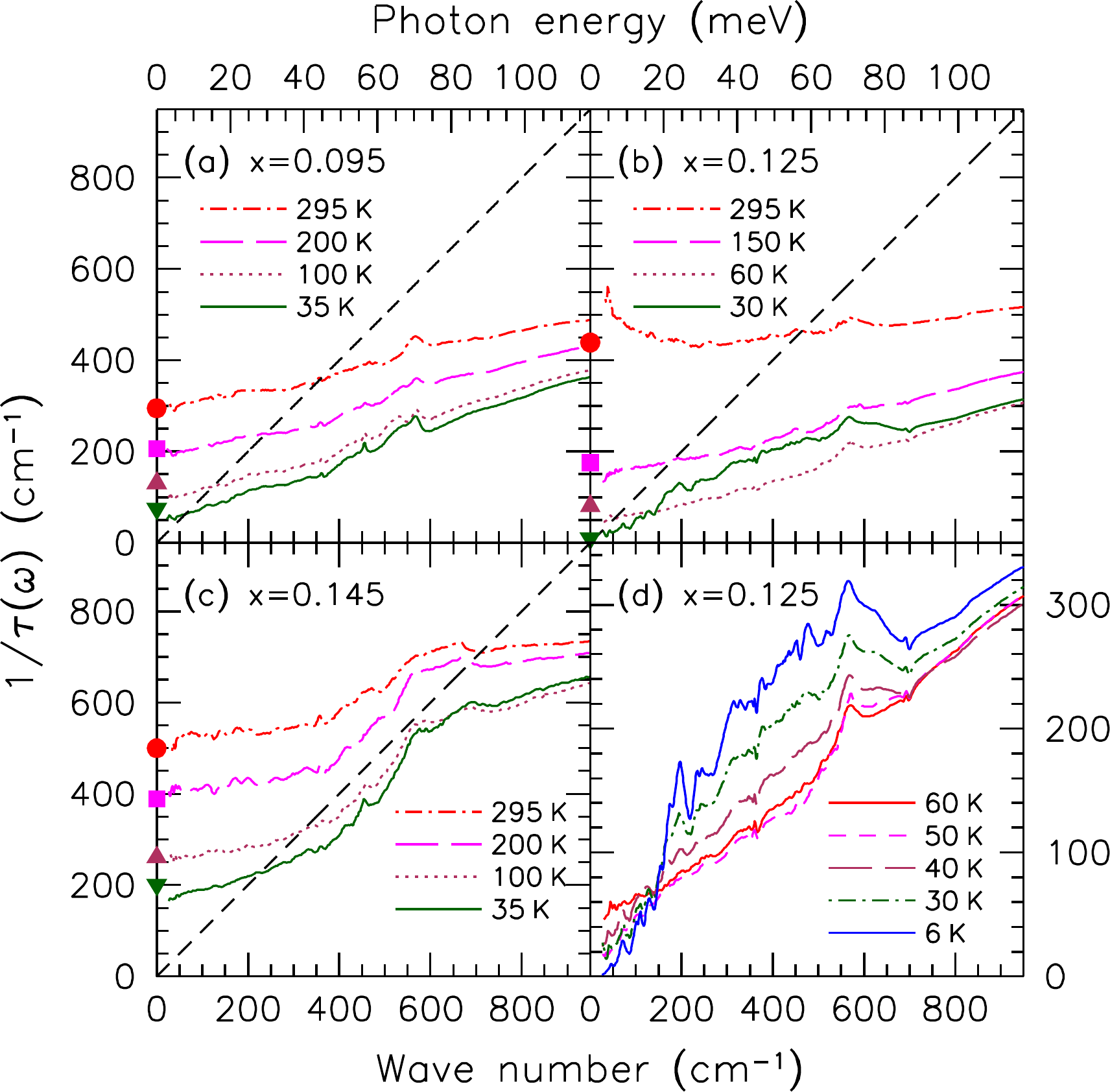}
\caption{The temperature dependence of the in-plane frequency-dependent
scattering rate in La$_{2-x}$Ba$_x$CuO$_4$ (with the phonons removed) in
the far-infrared region for (a) $x=0.095$, (b) the 1/8 phase, and
(c) $x=0.145$.
(d) Detailed temperature dependence of the scattering rate for
the 1/8 phase below 60~K.
The values for $1/\tau_D$ determined from the Drude fits are indicated
by the symbols at the origin. The dashed line indicates $1/\tau(\omega)
= \omega$.}
\label{fig:tau}
\end{figure}

The Drude model is a reasonable description of a non-interacting Fermi liquid;
however, the cuprates are either moderately or strongly correlated electron
systems.\cite{qazilbash09}  In this latter case, a more general description
of the Drude model is favored in which the scattering rate and the effective
mass are allowed to adopt a frequency dependence,\cite{allen77,puchkov96}
\begin{equation}
  {{1}\over{\tau(\omega)}} = {{\omega_p^2}\over{4\pi}} \,
  {\rm Re} \left[ {{1}\over{\tilde\sigma(\omega)}} \right]
\end{equation}
and
\begin{equation}
  {{m^\ast(\omega)}\over{m_e}} = {{\omega_p^2}\over{4\pi\omega}} \,
  {\rm Im} \left[ {{1}\over{\tilde\sigma(\omega)}} \right] ,
\end{equation}
where $m_e$ is the bare mass, $m^\ast(\omega)/m_e = 1+\lambda(\omega)$
and $\lambda(\omega)$ is a frequency-dependent electron-boson
coupling constant.  In this instance we set $\omega_p \equiv \omega_{p,D}$
and $\epsilon_\infty = 4$ (although the choice of $\epsilon_\infty$ has
little effect on the scattering rate or the effective mass in the
far-infrared region).  The infrared-active lattice modes observed in
the conductivity have been fit using Lorentzian oscillators superimposed
on a linear background and then removed from the complex dielectric
function.  The resulting temperature dependence of the in-plane
scattering rate $1/\tau(\omega)$ is shown in Fig.~\ref{fig:tau} for the
three different dopings studied in this work.  A simple Drude model
would produce a frequency-independent scattering rate; however,
despite the reasonable agreement with the Drude model, in
Figs.~\ref{fig:tau}(a), (b), and (c), $1/\tau(\omega)$ displays
an approximately linear frequency dependence.  We note that in the
$\omega\rightarrow 0$ limit the two models should yield the same
scattering rate, and indeed $1/\tau(\omega\rightarrow 0)
\simeq 1/\tau_D$.

%
%
The temperature dependence $1/\tau_D$ (shown on the plots and for
each doping) merits a brief discussion. For the three dopings examined,
$x=0.095$, 0.125, and 0.145, the values of the Drude scattering rate at
295~K increase steadily from $1/\tau_D = 295$~cm$^{-1}$ to 439~cm$^{-1}$
and 500~cm$^{-1}$, respectively, indicating either stronger electronic
correlations or that the increase in barium
concentration is accompanied by increasing out-of-plane cation
disorder, which may in turn lead to an increase in elastic scattering
in the copper-oxygen planes.  While $1/\tau_D$ decreases reasonably quickly
with decreasing temperature for $x=0.095$, this trend is not as pronounced
for $x=0.145$.  As previously noted, in the 1/8 crystal $1/\tau_D$ decreases
steadily with temperature, but for $T < T_{\rm so}$ the  reduction of the
scattering rate is dramatic, an effect that is shown in more detail
in Fig.~\ref{fig:tau}(d).
%
%
%

%
%
The frequency-dependence of the scattering rate contains a wealth
of information.  The dashed line in Fig.~\ref{fig:tau} denotes
$1/\tau(\omega) = \omega$; the region below this line is associated
with a Landau-Fermi liquid regime where the quasiparticles are well
defined.  In the materials examined, at high temperature, $1/\tau(\omega)
> \omega$ at low frequency, indicating that the conductivity in the
{\em a-b} planes is strongly dissipative, and that the quasiparticles
are not well defined.  Across the far-infrared region $1/\tau(\omega)$
decreases with temperature and in the $x=0.095$ material, just above
$T_c$ at 35~K, and in the 1/8 phase at low temperature ($T < T_{\rm co}$),
$1/\tau(\omega) \simeq \omega$, indicating well-defined quasiparticles
and coherent transport; this condition is never satisfied in the $x=0.145$
material. At low frequency, $1/\tau(\omega)$ decreases with decreasing
temperature; however, in the 1/8 phase this trend is reversed for
$T \lesssim T_{\rm so}$ above about 150~cm$^{-1}$, shown in Fig.~\ref{fig:tau}(d),
where the scattering rate increases dramatically.  This behavior is
consistent with the formation of a momentum-dependent gap.

%
%
It is interesting to note that, aside from some artifacts due to
the imperfect nature of the phonon subtraction, for the $x=0.095$
material in Fig.~\ref{fig:tau}(a) the frequency dependent scattering
rate is essentially featureless except for a small kink at about
45~meV (also observed in the $x=0.145$ material).  In many
high-temperature superconductors, $1/\tau(\omega)$ often has
sharp inflection points in the normal state;\cite{schachinger08} when
an inversion of the scattering rate is performed\cite{schachinger03,
dordevic05a} these structures correspond to peaks in the electron-boson
spectral function.\cite{hwang07a,hwang07b,hwang08a}  It has
been proposed that the peak location is proportional to the position
of the magnetic scattering resonance\cite{he01,he02} in polarized
neutron scattering measurements $\Omega_r \approx 5.4\,k_{\rm B} T_c$, and
more generally that it scales with the superconducting transition
temperature,\cite{yang09} $\Omega_r \approx 6.3\,k_{\rm B} T_c$.
In the underdoped cuprates the presence of a pseudogap on the Fermi
surface is expected to produce kinks in the scattering rate at
roughly $\Omega_r + \Delta_{\rm pg}$ and $\Omega_r + 2\Delta_{\rm pg}$.\cite{hwang08b}
In the $x=0.095$ ($T_c = 32$~K) material, taking the average,
$\Omega_r \simeq 16$~meV, while $\Delta_{\rm pg} \simeq 15$~meV
(Ref.~\onlinecite{valla06}) yielding $\simeq 31$ and 46~meV;
the larger value is very close to where a kink is observed in
$1/\tau(\omega)$ in Fig.~\ref{fig:tau}(a).  While it is tempting
to associate this kink with a magnetic resonance, the absence of
such a feature in the underdoped La$_{2-x}$Sr$_{x}$CuO$_4$
material\cite{chang07,chang08} suggests a different origin for
this feature.
In the more heavily doped $x=0.145$ ($T_c \simeq 24$~K) material,
$\Omega_r \simeq 12$~meV and $\Delta_{\rm pg} \simeq 10$~meV suggesting
features at $\simeq 22$ and 32~meV; however, no obvious structure in
the scattering rate is observed in Fig.~\ref{fig:tau}(c) at either of
these energies.


%
%

%
%
\subsubsection{Parameter scaling}
\begin{figure}[t]
\includegraphics[width=0.95\columnwidth]{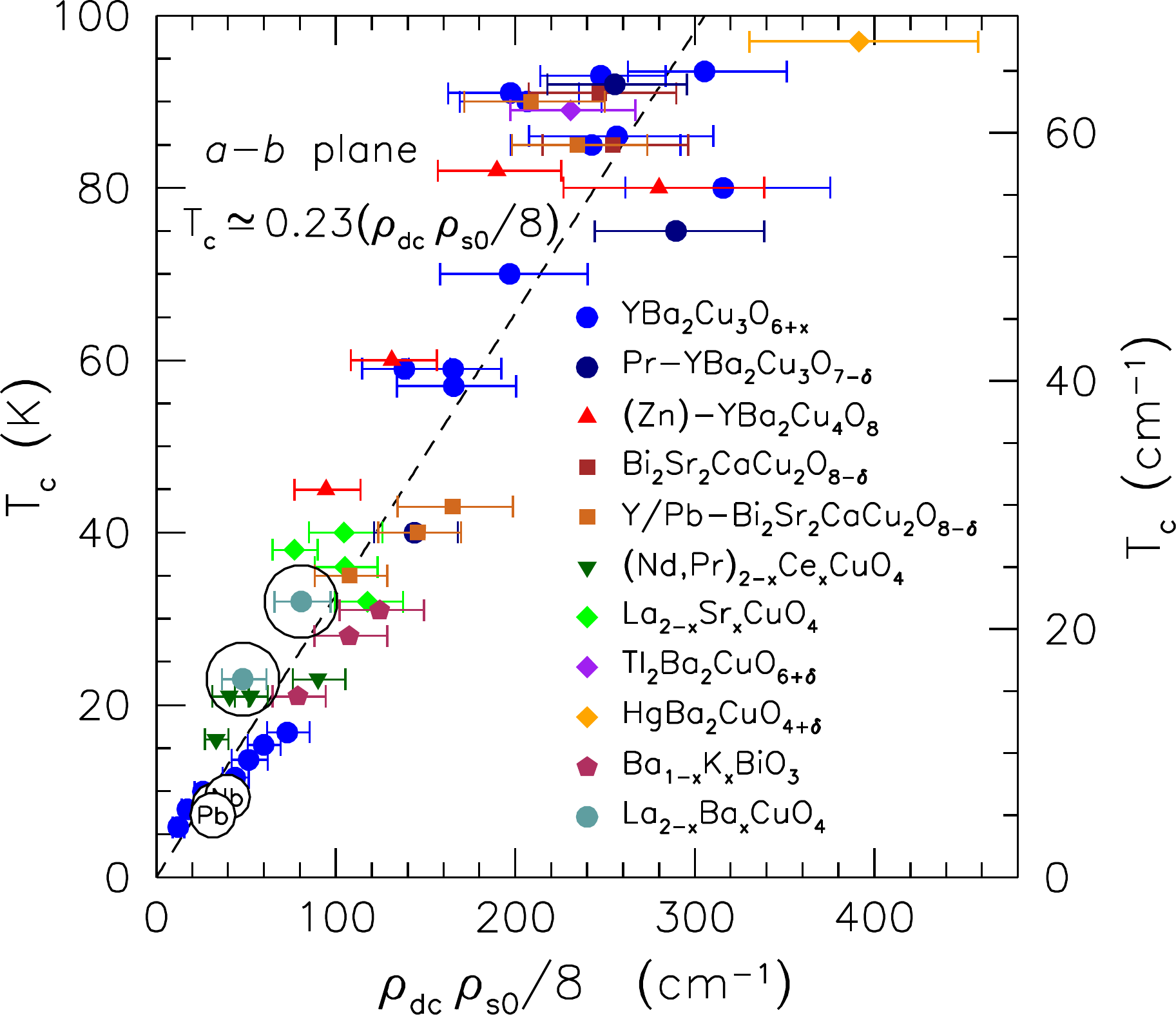}
\caption{The critical temperature $T_c$ versus the product of the dc
 resistivity measured just above $T_c$ and the spectral weight of the
 condensate,  $\rho_{dc}\,\rho_{s0}/8$, in the copper-oxygen planes
 for a variety of electron and hole-doped cuprates, compared with the
 {\em a-b} plane results for La$_{2-x}$Ba$_x$CuO$_4$ for $x=0.095$ and
 0.145 (highlighted within the circles).  The dashed line corresponds
 to the general result for the cuprates $T_c \simeq
 0.23\,(\rho_{dc}\,\rho_{s0}/8)$.}
\label{fig:scaling}
\end{figure}

It has been noted that the high-temperature superconductors obey the
scaling relation $\rho_{s0}/8 \simeq 4.4\,\sigma_{dc}\,T_c$, where
$\sigma_{dc}=\sigma_1(\omega\rightarrow 0)$ is the dc conductivity in the
normal state measured at $T \gtrsim T_c$, and the superfluid density $\rho_{s0}$ is
determined for $T \ll T_c$; this scaling relation is valid for both
the metallic {\em a-b} planes as well as along the poorly conducting {\em c}
axis, in both the electron and hole-doped cuprates.\cite{homes04,homes05a,homes05b,homes09}
In this representation, the scaling relation is valid over five orders
of magnitude and as such is usually shown as a log-log plot.
However, when only the {\em a-b} plane results are considered, the
results span less than two orders of magnitude and it is convenient
to rewrite this relation as
\begin{equation}
  T_c \simeq 0.23\,(\rho_{dc}\,\rho_{s0}/8),
\end{equation}
where $\rho_{dc} = 1/\sigma_{dc}$.  The scaling relation now has a similar
appearance to the well known Uemura relation,\cite{uemura89} $T_c \propto
\rho_{s0}$, and may be shown using a linear scale.  In Fig.~\ref{fig:scaling}
the values for $T_c$ are plotted against $\rho_{dc}\,\rho_{s0}/8$
determined in the {\em a-b} planes for a variety of single-layer and
double-layer cuprates.
The values for La$_{2-x}$Ba$_{x}$CuO$_4$ for $x=0.095$
($T_c=32$~K, $\sigma_{dc}=4100\pm{400}$~$\Omega^{-1}$cm$^{-1}$ just above
$T_c$, and $\rho_{s0}=12.6\pm{1}\times 10^6$~cm$^{-2}$) and $x=0.145$
($T_c=24$~K, $\sigma_{dc}= 4800\pm{400}$~$\Omega^{-1}$cm$^{-1}$
just above $T_c$, and $\rho_{s0}=8.8\pm{0.8}\times 10^6$~cm$^{-2}$)
are highlighted within the circles and fall on the general scaling line.
These points also lie quite close to the related La$_{2-x}$Sr$_{x}$CuO$_4$
materials with comparable doping.  This result is consistent with the
observation that there is strong dissipation in the normal
state.\cite{zaanen04,tallon06,lindner10,basov11b}
%
%

%
%
%
\subsection{{\em c} axis}
\subsubsection{Optical properties}
\begin{figure}[t]
\includegraphics[width=0.95\columnwidth]{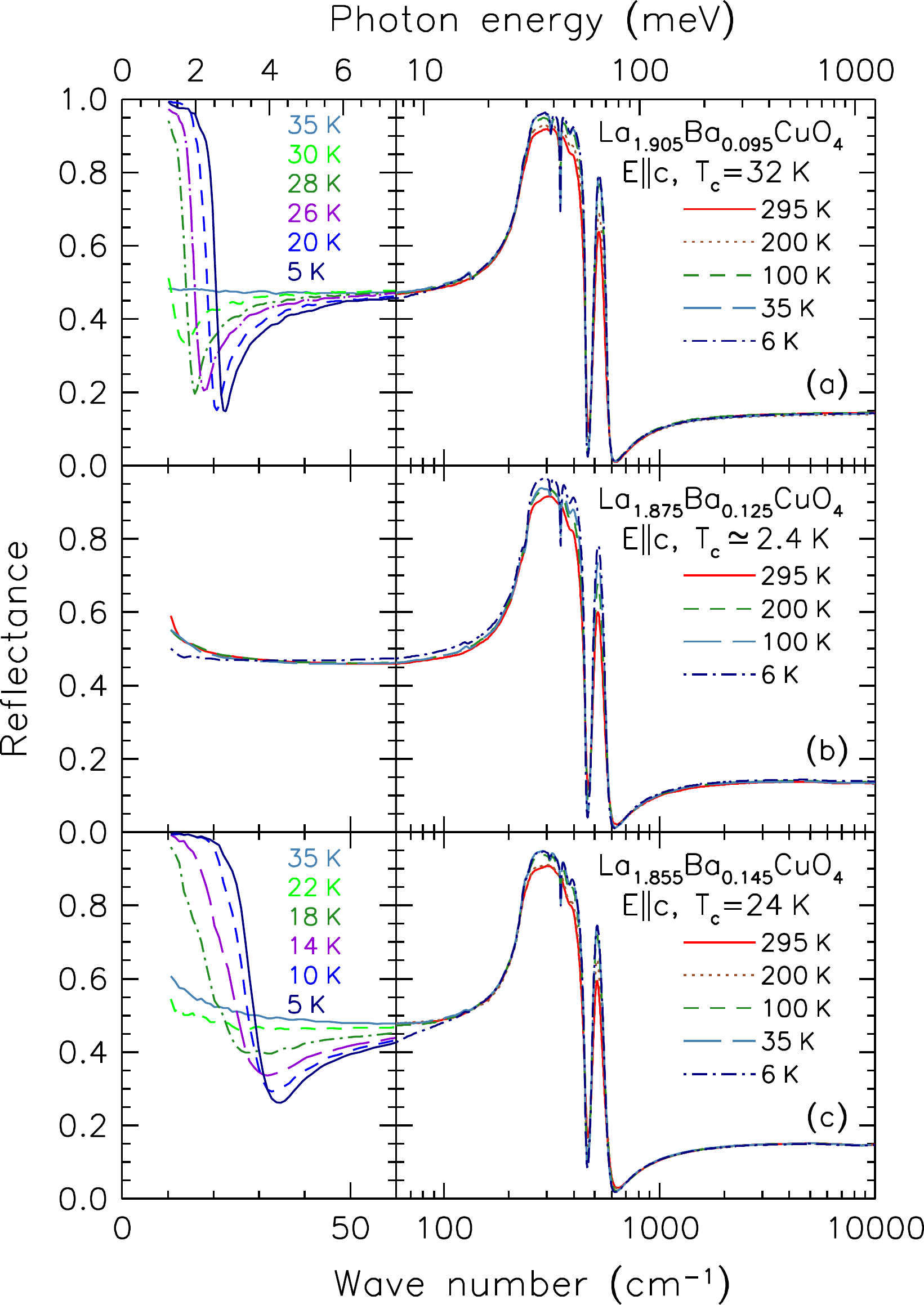}
\caption{The temperature dependence of the reflectance of La$_{2-x}$Ba$_x$CuO$_4$
for light polarized along the {\em c} axis for (a) $x=0.095$, (b) $x=0.125$, and
(c) $x=0.145$.  The wave number (photon energy) scale is linear up to 60~cm$^{-1}$;
above this it is logarithmic.}
\label{fig:creflec}
\end{figure}

The temperature dependence of the reflectance of La$_{2-x}$Ba$_x$CuO$_4$
for light polarized along the poorly-conducting {\em c} axis is shown in
Figs.~\ref{fig:creflec}(a), (b) and (c) for the $x=0.095$, 0.125 and 0.145
dopings, respectively.  The reflectance in this polarization is dramatically
different than what is observed in the metallic {\em a-b} planes
(Fig.~\ref{fig:reflec}); as a result of the near total absence of a
free-carrier contribution to the dielectric function, the reflectance
is dominated by the normally infrared-active {\em c} axis lattice modes.
In the normal state as $\omega\rightarrow 0$ the reflectance displays
a slight upturn, indicative of a weakly-conducting state; this is seen
most clearly in the $x=0.145$ material.
%
%
Below $T_c$ a sharp plasma edge is observed in the reflectance at
low frequency in Figs.~\ref{fig:creflec}(a) and (c) for the $x=0.095$
and 0.145 dopings, respectively.  The onset of the plasma edge
is extremely rapid in both materials.  The 1/8 phase material shows
no hint of a plasma edge in the reflectance; however, this is not
surprising as the bulk $T_c$ in this material is too low to be accessed
in our current experimental configuration.

The {\em c}-axis plasma edge in the reflectance is observed in many of
the cuprate superconductors,\cite{tamasaku92,homes93b,homes95,uchida96,
tajima97,dulic99,motohashi00,dordevic03,dordevic05b} and is attributed
to the Josephson coupling of the copper-oxygen planes resulting in the
formation  of a supercurrent and the onset of a bulk 3D superconducting
state.\cite{shibauchi94,basov94,radtke95}  It is
interesting to note that the Josephson plasma edge is considerably
broader in the more heavily-doped material; this observation is in
agreement with the result for heavily-doped La$_{2-x}$Sr$_x$CuO$_4$
where it is suggested that the smearing of the plasma edge is due to
spatial variations of the superconducting condensate.\cite{dordevic03}

%
%
\begin{table}[tb]
\caption{The vibrational parameters for oscillator fits to
the infrared-active phonon modes observed along the {\em c}
axis in La$_{1.905}$Ba$_{0.095}$CuO$_4$ at 295~K (HTT) and
at 35~K (LTO), where $\omega_j$, $\gamma_j$ and
$\Omega_j$  are the frequency, width and oscillator strength,
respectively.  The estimated errors are indicated in parenthesis.
All units are in cm$^{-1}$.}
\begin{ruledtabular}
\begin{tabular}{cccc}
 \multicolumn{4}{c}{$T=295$~K (HTT)} \\
 Mode     & $\omega_j$  & $\gamma_j$ & $\Omega_j$ \\
 $A_{2u}$ & 234.6 (0.2) &  29 (0.6)  & 1085 (10) \\
 $A_{2u}$ & 343.6 (0.1) & 9.2 (0.5)  &   61 (4) \\
 $A_{2u}$ & 381.0 (0.6) &  34 (3)    &   71 (10) \\
 $A_{2u}$ & 495.9 (0.1) &  25 (0.6)  &  312 (5)\\
          &             &            & \\
\multicolumn{4}{c}{$T=35$~K (LTO)} \\
 $B_{1u}$ & 132.7 (0.2) & 5.7 (0.9) &   55 (6) \\
 $B_{1u}$ & 235.4 (0.2) &  22 (0.6) & 1110 (10) \\
 $B_{1u}$ & 314.2 (0.1) & 7.1 (0.2) &   85 (2) \\
 $B_{1u}$ & 348.0 (0.1) & 4.6 (0.3) &  113 (6) \\
 $B_{1u}$ & 380.4 (0.9) &  39 (6)   &  109 (13) \\
 $B_{1u}$ & 495.8 (0.2) &  19 (0.6) &  338 (5) \\
\end{tabular}
\end{ruledtabular}
\label{tab:caxis}
\end{table}

\begin{figure}[t]
\includegraphics[width=0.95\columnwidth]{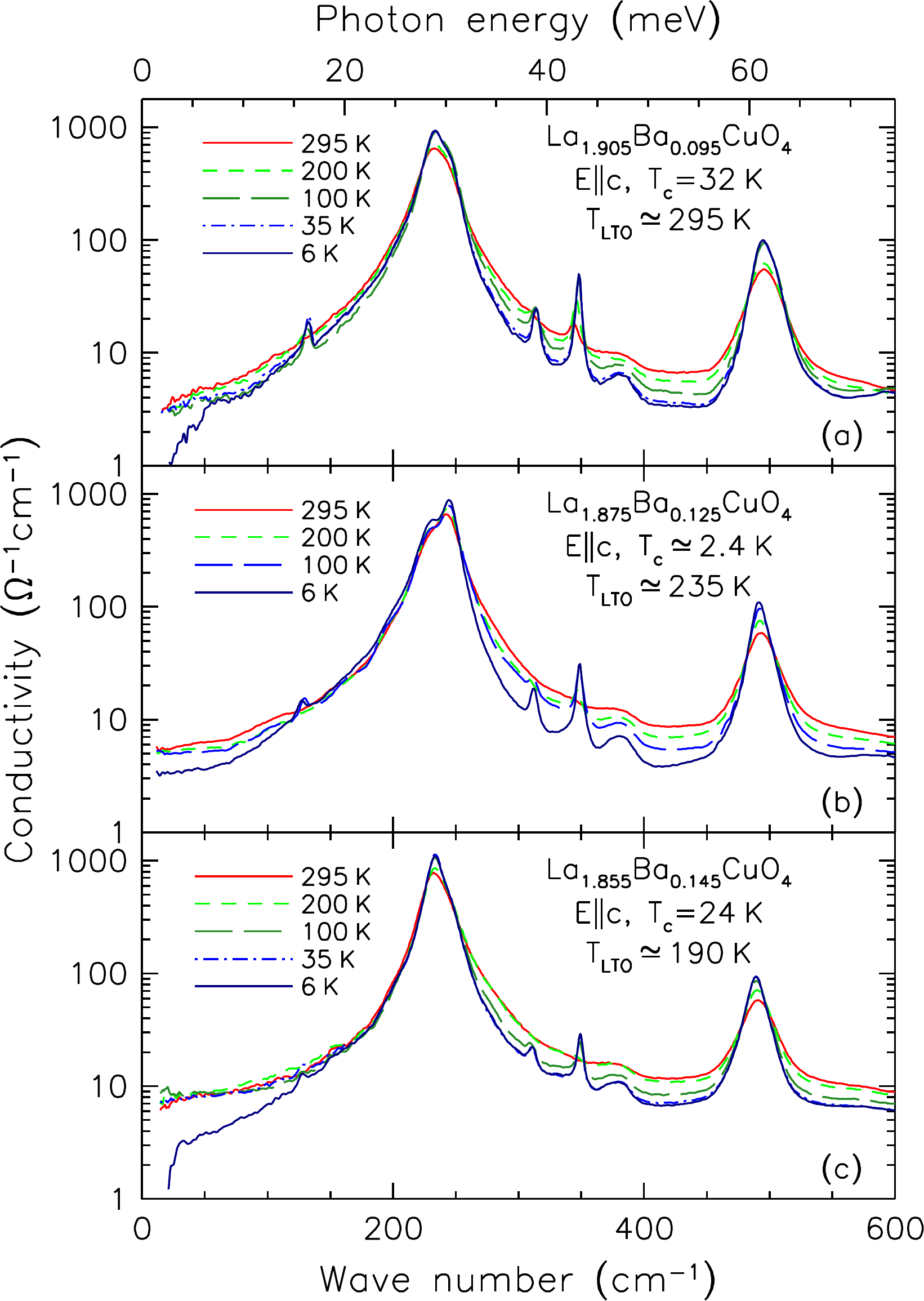}
\caption{The temperature dependence of the real part of the optical
conductivity of La$_{2-x}$Ba$_x$CuO$_4$ for light polarized along the
poorly-conducting {\em c} axis for (a) $x=0.095$, (b) $x=0.125$, and
(c) $x=0.145$. }
\label{fig:csigma}
\end{figure}

The behavior along the {\em c} axis is easier to understand when the
conductivity is considered, shown in Figs.~\ref{fig:csigma}(a), (b) and
(c) for $x=0.095$, 0.125 and 0.145, respectively.  In each case the
conductivity is dominated by a very strong infrared lattice mode at
$\simeq 235$~cm$^{-1}$ with another prominent mode at $\simeq 496$~cm$^{-1}$;
the conductivity is shown on a logarithmic scale to allow the emergence
of weaker modes, as well as the behavior of the electronic background,
to be examined.  The results of oscillator fits to the conductivity for
the $x=0.095$ material at 295 and 35~K are listed in Table~\ref{tab:caxis}.

In the HTT phase a total of four $A_{2u}$ modes are expected along
the {\em c} axis; despite $T_{\rm LTO}\simeq 305$~K in the $x=0.095$ material,
it is clear from the conductivity in Fig.~\ref{fig:csigma}(a) that only
four modes at $\simeq 235$, 344, 381 and 496~cm$^{-1}$ are present at room
temperature, indicating that the material is indeed in the HTT phase.
For $T < T_{\rm LTO}$ two new modes at $\simeq 133$ and 314~cm$^{-1}$ quickly
emerge, in agreement with the six $B_{1u}$ modes expected in the LTO phase.
It is unclear what effect, if any, the LTLO phase would have on the nature of
the {\em c} axis vibrations; however, at the lowest measured temperature
there is a slight asymmetry in the strong 235~cm$^{-1}$ mode, suggesting a
possible further reduction of symmetry.
The vibrational properties in the $x=0.125$ and 0.145 materials are similar.
At room temperature both of these materials are in the HTT phase and four
modes are observed; two new modes emerge below the respective values for
$T_{\rm LTO}$.  In the 1/8 phase, at $\sim 6$~K this material is expected
to be in an LTT phase and a total of seven $A_{2u}$ modes are expected to
be active along the {\em c} axis; the shoulder in the strong 235~cm$^{-1}$
mode suggests that all seven $A_{2u}$ modes may be observed.
The picture in the more heavily doped $x=0.145$ material is less clear.  While
six modes are observed at low temperature as expected for the LTT phase,
the modes in this material have broadened and it is no longer obvious if there
is any asymmetry in the 235~cm$^{-1}$ mode at low temperature.  This may be a
reflection of the somewhat ambiguous nature of the low-temperature structure
(LTT$+$LTLO) for this doping.\cite{hucker11}
%
%
%

%
%
The conductivity along the {\em c} axis is orders of magnitude
lower than what is observed in the metallic {\em a-b} planes. At
room temperature in the $x=0.095$ material $\sigma_{dc} \equiv
\sigma_1(\omega \rightarrow 0) \simeq 3$~$\Omega^{-1}$cm$^{-1}$,
increasing to $\simeq 6$~$\Omega^{-1}$cm$^{-1}$ in the 1/8 phase
[both values are comparable with the {\em c} axis resistivity
shown in Figs.~\ref{fig:trans}(a) and (b)], and
$\simeq 8$~$\Omega^{-1}$cm$^{-1}$ in the $x=0.145$ material.
The normal-state transport along the {\em c} axis is attributed
to thermally-activated hopping, which is consistent with the
observation of an electronic background that decreases (weakly) with
temperature.  In addition, some of this decrease in the optical
conductivity may also be due to a narrowing of vibrational features.

Despite the small value for the background conductivity, a suppression
of the low-frequency conductivity due to the formation of a condensate
is observed below $T_c$ in the $x=0.095$ and 0.145 materials.  The
decrease in the low-frequency conductivity in the 1/8 phase below $T_{\rm co}$
is likely due to the decoupling of the planes due to the formation of
charge-stripe order;\cite{berg07,wen10} however, the dramatic decrease
in the {\em c}-axis resistivity below $T_{\rm so}$ [Fig.~\ref{fig:trans}(b)]
is not  reflected in the optical properties as no commensurate increase in
the conductivity is observed down to the lowest measured frequency.  This
suggests that the decrease in the low-frequency spectral weight at low
temperature is associated with an increase in the optical conductivity
at very low energies (possibly microwave or radio frequencies), and is
likely associated with the onset below $T_{\rm so}$ of superconducting
correlations in the planes.\cite{tranquada08}
In the $x=0.095$ and 0.145 compounds the energy scale for the loss
of spectral weight is roughly 80~cm$^{-1}$ (10~meV) and 140~cm$^{-1}$
(18~meV), respectively.  It is interesting to note that these energy scales
correspond to $\Delta_0$, rather than $2\Delta_0$ observed in the {\em a-b}
planes.\cite{valla06}  This type of behavior has been previously
observed in the cuprates;\cite{basov96} however, it remains unexplained.

%
%
\begin{figure}[t]
\includegraphics[width=0.95\columnwidth]{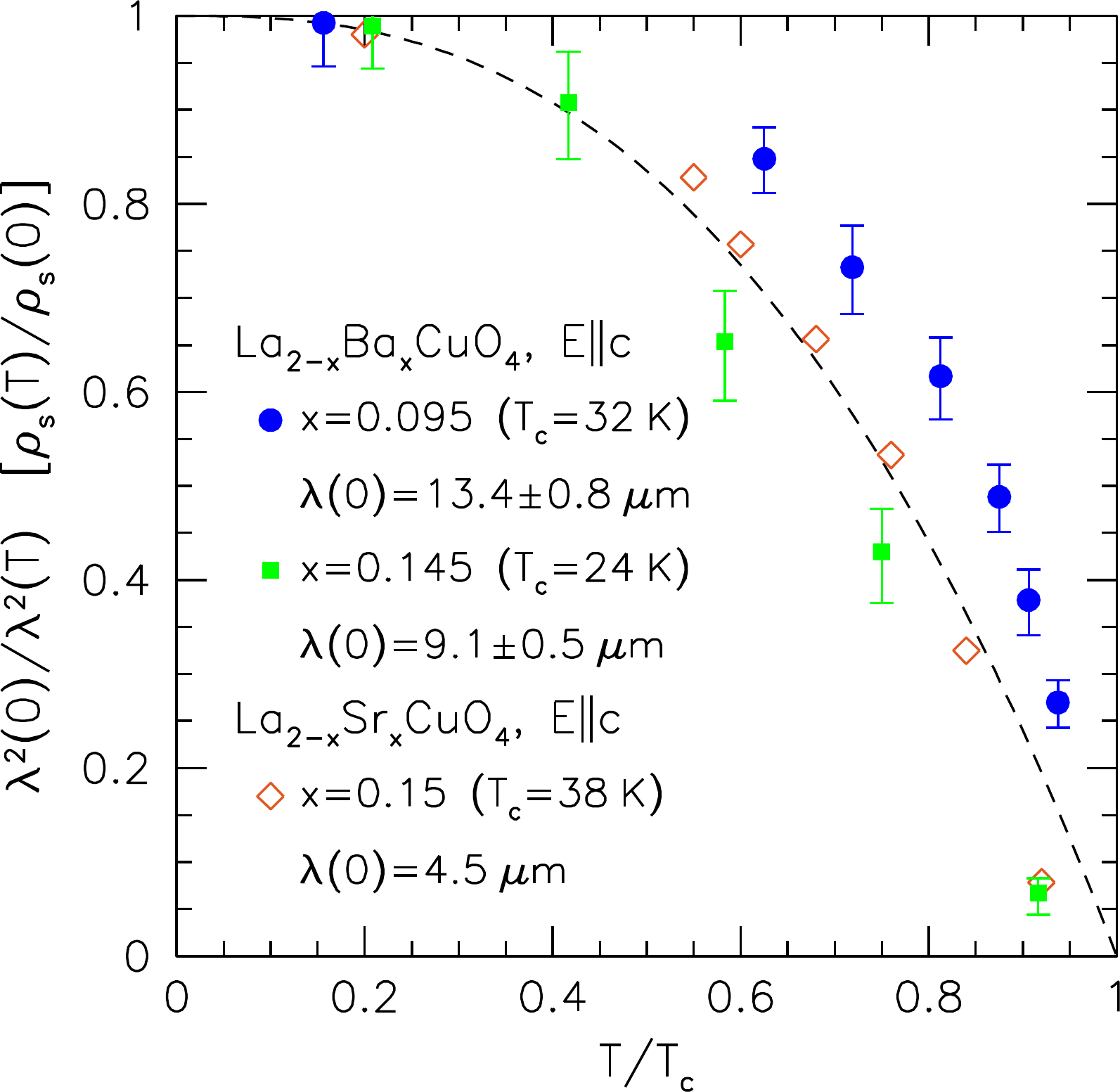}
\caption{The temperature dependence of the superfluid density along the {\em c}
axis normalized to an extrapolated zero-temperature value versus the reduced
temperature, $t=T/T_c$, in La$_{2-x}$Ba$_x$CuO$_4$ for $x=0.095$ ({\CIRCLE})
and 0.145 ($\blacksquare$) and optimally-doped La$_{1.85}$Sr$_{0.15}$CuO$_4$
($\diamondsuit$) (Ref.~\onlinecite{dordevic05b}). The dashed line is a guide
to the eye.}
\label{fig:lambda}
\end{figure}
%

%
%
\subsubsection{Penetration depth}
While it is possible to determine $\omega_{p,S}$ from the Ferrell-Glover-Tinkham
sum rule, the small values of the conductivity and the proximity of strong lattice
vibrations can lead to significant uncertainties.  Under these circumstances a
more reliable approach is to examine the low-frequency response of the real
part of the dielectric function, $\epsilon_1(\omega) = \epsilon_\infty -
\omega_{p,S}^2/\omega^2$.  For $T \ll T_c$ this method yields $\omega_{p,S}
= 119\pm{10}$~cm$^{-1}$ for $x=0.095$, and $\omega_{p,S}=174\pm{14}$~cm$^{-1}$
for $x=0.145$, which corresponds to effective penetration depths of
approximately 13.4 and 9.1~$\mu$m, respectively.
It is also possible to track the temperature dependence of the superfluid
density using this method.  The temperature dependence of the superfluid
density for the $x=0.095$ and 0.145 samples, expressed here as the ratio
value of the square of the penetration depths $\lambda^2(0)/\lambda^2(T)$,
is plotted against the reduced temperature $t=T/T_c$  and compared with
optimally-doped La$_{1.85}$Sr$_{0.15}$CuO$_4$, shown in Fig.~\ref{fig:lambda}.
In a comprehensive study of the evolution of the {\em c}-axis superfluid
density in La$_{2-x}$Sr$_x$CuO$_4$ for a wide variety of dopings,\cite{dordevic05b}
it may be noted that the superfluid density develops most rapidly for the
optimally-doped material.  It is therefore surprising that in underdoped
La$_{1.905}$Ba$_{0.095}$CuO$_4$ the superfluid density develops extremely
quickly just below $T_c$ before slowing below $T_c/2$; in this regard it
is reminiscent of the behavior of a system with an isotropic {\em s}-wave
energy gap,\cite{prohammer91} while the more heavily doped $x=0.145$
sample develops relatively slowly and is similar to what is observed
in the copper-oxygen planes in a {\em d}-wave system.\cite{zhang94}

%
%
We attribute this difference in behavior to the presence of a pseudogap in
the underdoped materials where the antinodal region is gapped well above $T_c$.
If the {\em c}-axis transport is dominated by the antinodal zone boundary
region of the Fermi surface,\cite{chakravarty93,xiang96,ioffe99} then when the
Fermi pocket is gapped below $T_c$ the {\em c}-axis response will be dominated
by the gap maximum $\Delta_0$ well away from the nodal region, and thus will
not be sensitive to the {\em d}-wave nature of the system, instead resembling a
more isotropic system.\cite{homes05b} The formation of charge order in the
$x=0.095$ material below $\simeq 25$~K may frustrate the Josephson coupling
between the planes,\cite{wen10} leading to the observed slowing of the formation
of the condensate below roughly $T_c/2$.
On the other hand, the pseudogap is generally not present in materials close
to optimal doping; below $T_c$ the entire Fermi surface is gapped.  This suggests
that in a optimally-doped material the {\em c}-axis response will be more
sensitive to the {\em d}-wave nature of the energy-gap and therefore evolves
more slowly, similar to the response observed in the copper-oxygen planes.\cite{hosseini98}

%
%
\begin{figure}[t]
\includegraphics[width=0.95\columnwidth]{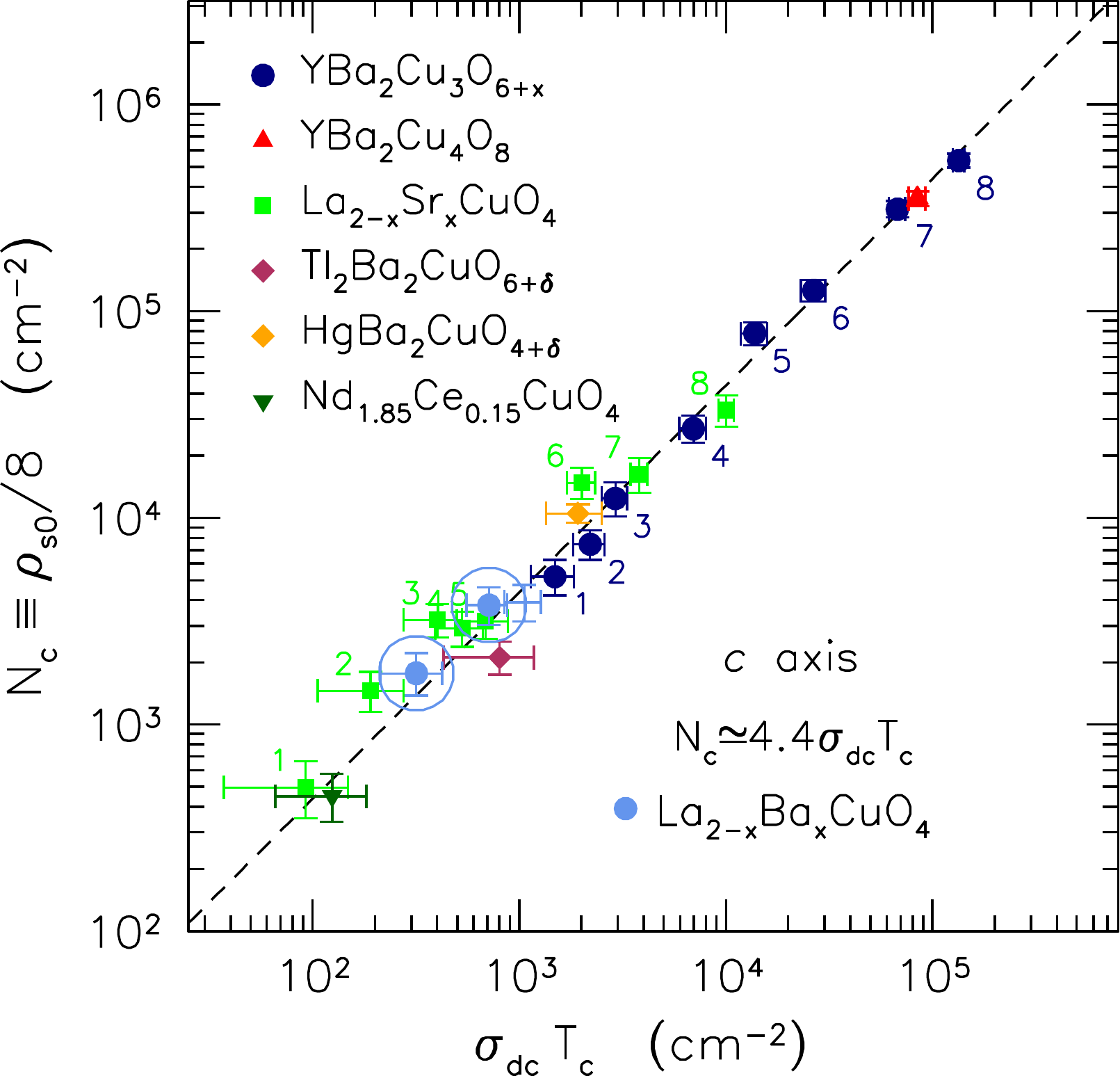}
\caption{The log-log plot of the spectral weight of the superfluid
 density $N_c \equiv \rho_{s0}/8$ vs $\sigma_{dc}\,T_c$ along the
 {\em c} axis for a variety of electron and hole-doped cuprates
 compared with La$_{2-x}$Ba$_x$CuO$_4$ for $x=0.095$
 and 0.145 (highlighted within the circles).  The superscripts next
 to the symbols for La$_{2-x}$Sr$_x$CuO$_4$ and YBa$_2$Cu$_3$O$_{6+y}$
 refer to different chemical dopings.\cite{homes05a}  The dashed line
 corresponds to the general result for the cuprates $\rho_{s0}/8
 \simeq 4.4 \sigma_{dc} T_c$.}
\label{fig:cscale}
\end{figure}
%

%
%
\subsubsection{Parameter scaling}
It was previously noted that the scaling  relation $\rho_{s0}/8
\simeq 4.4\,\sigma_{dc}T_c$ is valid not only in the {\em a-b}
planes, but also along the {\em c} axis.\cite{homes04} While it is
possible to plot the {\em c} axis data in the form of $T_c$ versus
$\rho_{dc}\,\rho_{s0}/8$, the highly anisotropic nature of these
materials results in conductivities (resistivities) along the {\em c}
axis that are orders of magnitude smaller (larger) than those
observed in the {\em a-b} planes, resulting in a significant
uncertainty in $\sigma_{dc}=\sigma_1(\omega\rightarrow 0)$ and
the corresponding value for $\rho_{dc}=1/\sigma_{dc}$.  In
addition, the scaling behavior of $\rho_{s0}$ emerges naturally
if the superconductivity along the {\em c} axis is considered to
originate from Josephson coupling.\cite{homes05b}

The {\em c} axis values for the spectral weight of the condensate
$\rho_{s0}/8$ are plotted as a function of $\sigma_{dc}\,T_c$ in
Fig.~\ref{fig:cscale} for a variety of single-layer and double-layer
cuprates.  The dashed line is the best fit to the data,
$\rho_{s0}/8 \simeq 4.4\,\sigma_{dc}\,T_c$.
The {\em c}-axis values for La$_{2-x}$Ba$_{x}$CuO$_4$ for $x=0.095$
($T_c=32$~K, $\sigma_{dc}=3\pm{1}$~$\Omega^{-1}$cm$^{-1}$ just above
$T_c$, and $\rho_{s0}=1.4\pm{0.2}\times 10^4$~cm$^{-2}$) and 0.145
($T_c=24$~K, $\sigma_{dc}= 9\pm{2}$~$\Omega^{-1}$cm$^{-1}$
just above $T_c$, and $\rho_{s0}=3.0\pm{0.3}\times 10^4$~cm$^{-2}$)
are highlighted in the circles.  These values fall on the general
scaling line and once again are quite close to the related
La$_{2-x}$Sr$_{x}$CuO$_4$ materials with comparable doping.

%
%
%
\subsection{The anomalous 1/8 phase}
The behavior of the optical conductivity at low temperature in the $x=1/8$
phase is fundamentally different than the other two dopings, where a
response that is consistent with the opening of a superconducting {\em d}-wave
energy gap is observed.  The transfer of spectral weight from low to high
frequency that develops close to or below $T_{\rm so}$ is extremely unusual
for a hole-doped cuprate and has the characteristics of an anisotropic
transport gap on the Fermi surface due to the formation of charge or spin
order.
%
%
%
A partial gapping of the Fermi surface has been observed in other 2D materials
with incommensurate charge-density waves (CDW's).\cite{hess90,wang90,mcconnell98,dordevic01}
Unlike one-dimensional systems, where the modulation wave vector can perfectly
span the Fermi surface and the CDW almost always leads to the complete gapping
of the Fermi surface, in 2D perfect nesting is not possible\cite{gruner} and as
a result the CDW gaps only a portion of the Fermi surface.
%
%
Fortunately, ARPES measurements have been performed on this composition,\cite{valla06,he09}
so that we can analyze nesting of the Fermi surface by the measured modulation wave
vectors.\cite{hucker11}  The charge-order wave vector is approximately equal to $4k_{\rm F}$
in the antinodal region.  There is certainly a gap on that portion of the Fermi surface, but
it is already there as a pseudogap before the charge-stripe order appears.\cite{valla06,he09}
Therefore, the temperature dependence of the optical spectral weight transfer suggests a connection
with the spin-stripe order for which the wave vector would nest parts of the Fermi arc.
The ARPES measurements indicate that the gapping of the Fermi arc below $T_{\rm so}$ is
indistinguishable from the $d$-wave-like gap found in superconducting samples.\cite{valla06,he09}

%
%
%
%
%
%
While the conventional density-wave picture is a possible candidate to explain the
observed redistribution of spectral weight at low temperature, another possibility
is that this shift is associated with superconducting correlations in this
material.\cite{tranquada08}
We are not able to access the bulk superconducting state in the $x=1/8$ material; however,
the previously discussed scaling relation\cite{homes04} indicates that because of
the dramatically reduced value for the bulk superconductivity in this material
($T_c \simeq 2.4$~K) the Josephson plasma frequency along the {\em c} axis is
estimated to be $\omega_{p,S} \simeq 17$~cm$^{-1}$. The Josephson plasma edge
in the reflectance typically occurs in a region close to the renormalized
superconducting plasma frequency $\tilde\omega_{p,S} =  \omega_{p,S}/
\sqrt{\epsilon_{\rm FIR}}$.  Given the experimentally-determined value
of $\epsilon_{1,c} \simeq 30$ at 50~cm$^{-1}$, this yields a value of
$\tilde\omega_{p,S} \simeq 3$~cm$^{-1}$, below our ability to measure
using current optical techniques.
The strongly reduced value for the {\em c}-axis Jopsephson plasma edge for
$x=1/8$ is consistent with the disappearance of this feature in
La$_{1.85-y}$Nd$_y$Sr$_{0.15}$CuO$_4$, which occurs when the increasing
Nd concentration causes the structure to change from LTO to LTT, resulting
in stripe order.\cite{tajima01,schafgans10}  To explain the abrupt change
in the Josephson coupling in the latter case, despite the much weaker
reduction in $T_c$, Himeda {\it et al.}\cite{hime02} proposed a pair-density-wave
(PDW) superconducting state involving local $d$-wave symmetry but modulated by a
sinusoidal envelope function; the envelope function has the same period as the spin
modulation, but its extrema are aligned with the maxima of the charge modulation.
The rotation of the stripe (and PDW) orientation by $90^\circ$ from layer to layer
leads to a frustration of the interlayer Josephson coupling.  This concept was
rediscovered\cite{berg07,berg09b} after the observation of quasi-two-dimensional
superconductivity by transport measurements\cite{li07} on La$_{2-x}$Ba$_x$CuO$_4$
with $x=1/8$.

\begin{figure}[t]
\includegraphics[width=0.95\columnwidth]{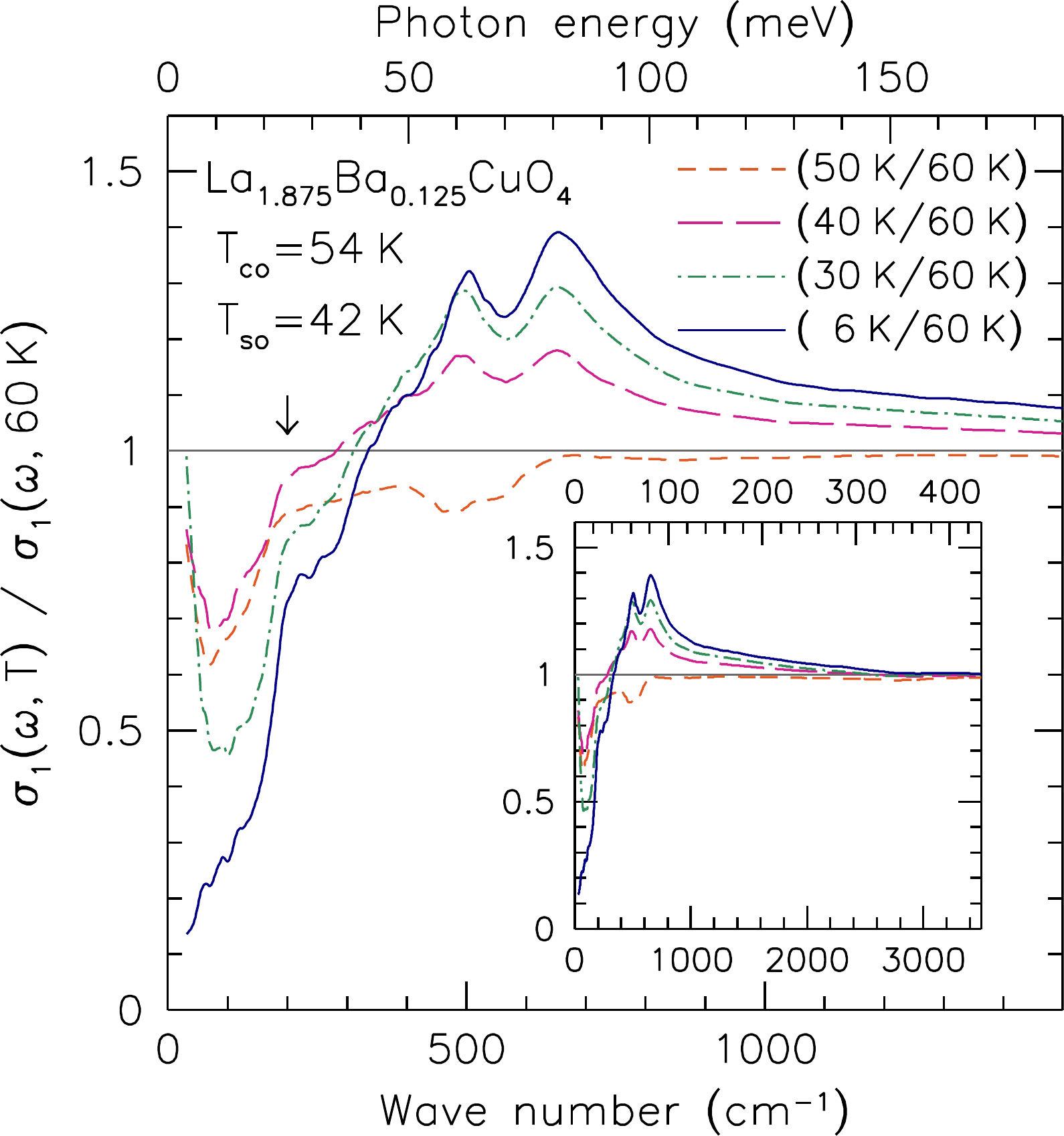}
\caption{The temperature dependence of the smoothed ratio of the
real part of the optical conductivity with the 60~K data. The phonons
and antiresonances have been subtracted.  The arrow denotes a
kink attributed to the formation of a gap on the Fermi arc
below $T_{\rm so}$.
Inset: The ratio over a wide frequency range.}
\label{fig:ratio}
\end{figure}

The PDW proposal also has implications for the in-plane conductivity.  Given a
sinusoidally-modulated pair wave function, the superfluid density can be written
as
\begin{equation}
\rho_s({\bf r}) = 0.5\rho_s [1 + \cos(2{\bf q}\cdot{\bf r}+\phi)],
\end{equation}
where ${\bf q}=(\frac18,0,0)$ [or $(0,\frac18,0)$] is the ordering wave
vector (in reciprocal lattice units based on the HTT phase), and $\phi$ is
the phase of the modulation with respect to the crystal lattice.  Barabash
{\it et al.}\cite{bara00} showed that inhomogeneity of the superfluid
density will cause some of the conductivity that would normally be at
$\omega=0$ to shift to finite $\omega$.  Orenstein\cite{oren03} went
further, and showed that sinusoidal modulation of $\rho_s({\bf r})$
shifts conductivity to a peak at the renormalized superconducting plasma
frequency $\tilde\omega_{p,S} \approx \omega_{p,S}/\sqrt{\epsilon_\infty}$.
From Table~\ref{tab:props} we extrapolate an in-plane value of $\omega_{p,S} \simeq
3250$~cm$^{-1}$ for $x=1/8$.  Using the experimentally-determined value of
$\epsilon_{1,a} \simeq 4$ at 2~eV for $\epsilon_\infty$ [there
is little or no temperature dependence of $\epsilon_{1,a}(\omega)$ in this
frequency region], then we estimate the in-plane $\tilde\omega_{p,S}
\simeq 1600$~cm$^{-1}$.
In considering Fig.~\ref{fig:sigma3}(b), we already noted that some spectral weight
is shifted from low frequency to above 300 cm$^{-1}$.  To emphasize this behavior,
Fig.~\ref{fig:ratio} shows the ratio of $\sigma_1(\omega,T)$ to that at $T=60$~K.
The enhancement of $\sigma_1$ at $\sim80$~meV is clearly present at 40~K and has
doubled by 6~K; however, there is no enhancement at 50~K, which is already below
$T_{\rm co}$.  The development of the 80-meV feature on cooling correlates with
the drop in in-plane resistivity.

To compare with Orenstein's analysis,\cite{oren03} we have to allow for the
fact that we have superconducting correlations but not long-range order for
40~K$\null\gtrsim T > 16$~K.  As a consequence, there should be a finite scattering
rate associated with the ``superconducting'' features.  The small but finite
width of the $\omega=0$ (``Drude'') peak for 16~K$\null<T \lesssim 50$~K is consistent
with this expectation.  The development of the finite-frequency peak is qualitatively
compatible with Orenstein's prediction for a modulated superconductor, although it
appears at a frequency somewhat lower than $\tilde\omega_{p,S}$.  The shifted
spectral weight appears to be pushed above $2\Delta_0$, where ARPES
studies\cite{valla06,he09} indicate $2\Delta_0\approx 40$~meV in the
antinodal region.

There are some experimental results that complicate this picture.  The ARPES
studies\cite{valla06,he09} indicate a $d$-wave-like gap along the near-nodal arc
for $T \lesssim 40$~K, whereas the PDW state is predicted to have a gapless
near-nodal arc.\cite{baru08}  The $d$-wave-like gap suggests the presence of a
more uniform superconducting component that develops below $\sim 40$~K.  Such a
uniform component might be necessary to get bulk superconductivity at low
temperature, as the correlation length of the PDW order is limited by that
of the stripe order, which is restricted to several hundred \AA.\cite{hucker11,wilk11}
Of course, if a uniform component is present, then one returns to the challenge
of explaining why the interlayer Josephson coupling is frustrated.

A different complication is that a substantial narrowing of the zero-frequency peak
is already apparent at 50~K, where we have no enhancement at $\sim 80$~meV.  At that
temperature, we have charge stripe order but not spin order, and the spin order may
be necessary to pin the phase of $\rho_s({\bf r})$ with respect to the lattice.
Perhaps at 50~K, in the absence of pinning, the PDW correlations start to show up
at low frequency but not at higher frequency.  In any case, the high-frequency peak
remains at 6~K, below the point where 2D superconducting order develops,\cite{tranquada08}
so that the modulated $\rho_s$ should still be present there.

%
%
%
\section{Conclusions}
The temperature dependence of the optical properties of La$_{2-x}$Ba$_x$CuO$_4$
for light polarized in the {\em a-b} planes and along the poorly-conducting
{\em c} axis has been determined for $x=0.095$, 0.125 and 0.145.  The $x=0.095$
and 0.145 samples are superconducting, with $T_c = 32$ and $\simeq 24$~K,
respectively.  The in-plane optical conductivity can be characterized as a
Drude-like response that narrows with decreasing temperature, superimposed
on a flat, incoherent background; below $T_c$ missing low-frequency spectral
weight is the hallmark of the formation of a superconducting condensate.
Interestingly, in both cases $\omega_{p,S}^2/\omega_{p,D}^2 < 1$, indicating
that only a fraction of the normal-state carriers condense; this is consistent
with the observation of strong dissipation in the normal state.

The optical properties along the poorly-conducting {\em c} axis in the
$x=0.095$ and 0.145 materials are insulating in character, dominated
by the normally infrared-active phonons with only a weak electronic
background; new lattice modes appear at the HTT$\rightarrow$LTO transition.
Below $T_c$ a sharp plasma edge is observed in the {\em c}-axis reflectance;
this feature is attributed to the formation of a supercurrent along the
{\em c} axis due to Josephson coupling between the copper-oxygen planes.
Despite the dramatic reduction in the {\em c}-axis resistivity below
$T_{\rm so}$ in the 1/8 material, no Josephson plasma edge or hint of
metallic behavior is observed.
In the superconducting materials, the spectral weight of the condensate
for the {\em a-b} plane and the {\em c} axis fall on the universal
scaling line $\rho_{s0}/8 \simeq 4.4\,\sigma_{dc}\,T_c$.

The behavior of the anomalous $x=1/8$ phase in which the superconductivity is
dramatically weakened ($T_c \simeq 2.4$~K) stands out in stark contrast.  The
in-plane optical properties are similar to the other two dopings at high
temperature; however, close to or below $T_{\rm so}$ there is an unusual
suppression of the reflectance over much of the infrared region and a transfer
of spectral weight from low frequency to energies above $\simeq 40$~meV,
indicating the formation of a momentum-dependent energy gap.  Below $T_{\rm so}$
the free-carrier response continues to narrow dramatically with decreasing
temperature, to the extent that at the lowest measured temperature the
spectral weight associated with the free carriers can no longer be observed.
This response mimics the missing spectral weight associated with the formation
of a condensate; however, we are currently unable to distinguish a
momentum-dependent CDW gap from the similar behavior expected from a
PDW state.

%
%
\begin{acknowledgements}
We would like to thank A. Akrap, A. Auerbach, D. N. Basov,
D. A. Crandles, S. V. Dordevic, J. Hwang, S. A. Kivelson,
M. Reedyk, T. Timusk, and N. L. Wang for useful discussions,
and M. G. Rechner for careful reading of this manuscript.
Research supported by the U.S. Department of Energy, Office of
Basic Energy Sciences, Division of Materials Sciences and Engineering
under Contract No. DE-AC02-98CH10886.
\end{acknowledgements}
%

%
%
%

\begin{thebibliography}{146}
\bibitem [{fit()}]{fit}%
  \BibitemOpen
  \href@noop {} {}\bibinfo {note} {To allow for the data to be fit over the
  entire frequency range, two high-frequency oscillators have also been
  included at $\simeq 13\,000$ and $21\,000$~cm$^{-1}$.}\BibitemShut {Stop}%
\end{thebibliography}

\providecommand{\noopsort}[1]{}\providecommand{\singleletter}[1]{#1}

\end{document}